\documentclass[]{jfm}

\usepackage{graphicx}
\usepackage{newtxtext}
\usepackage{newtxmath}
\usepackage{natbib}
\usepackage{hyperref}
\usepackage{placeins}
\hypersetup{
    colorlinks = true,
    urlcolor   = blue,
    citecolor  = black,
}

\newcommand{\RomanNumeralCaps}[1]
\linenumbers

\shorttitle{Equilibria and stability of plates}
\shortauthor{O. Pomerenk and L. Ristroph}

\title{Aerodynamic equilibria and flight stability of plates at intermediate Reynolds numbers}

\author{Olivia Pomerenk\aff{1} \and Leif Ristroph\aff{1}\corresp{\email{ristroph@cims.nyu.edu}}}
\affiliation{\aff{1}Courant Institute of Mathematical Sciences, Applied Math Lab, New York University, New York, NY 10012 }

\begin{document}

\maketitle

\begin{abstract}

The passive flight of a thin wing or plate is an archetypal problem in flow-structure interactions at intermediate Reynolds numbers. This seemingly simple aerodynamic system displays an impressive variety of steady and unsteady motions that are familiar from fluttering leaves, tumbling seeds and gliding paper planes. Here, we explore the space of flight behaviors using a nonlinear dynamical model rooted in a quasi-steady description of the fluid forces. Efficient characterization is achieved by identification of the key dimensionless parameters, assessment of the steady equilibrium states, and linear analysis of their stability. The structure and organization of the stable and unstable flight equilibria proves to be complex, and seemingly related factors such as mass and buoyancy-corrected weight play distinct roles in determining the eventual flight patterns. The nonlinear model successfully reproduces previously documented unsteady states such as fluttering and tumbling while also predicting new types of motions, and the linear analysis accurately accounts for the stability of steady states such as gliding and diving. While the conditions for dynamic stability seem to lack tidy formulas that apply universally, we identify relations that hold in certain regimes and which offer mechanistic interpretations. The generality of the model and the richness of its solution space suggest implications for small-scale aerodynamics and related applications in biological and robotic flight. 
\end{abstract}

\begin{keywords}
aerodynamic modeling $|$ unsteady aerodynamics $|$ passive flight $|$ flight dynamics $|$ flow-structure interaction
\end{keywords}

\section{Introduction}

Thin structures falling through fluids exhibit a wide variety of unsteady and steady motions such as fluttering, tumbling, and gliding. Such passive flight systems are canonical areas of study for aerodynamics at intermediate Reynolds numbers ($\Rey$) and the interactions of bodies with unsteady flows. Attempts to identify and categorize flight behaviors and understand their physical origins date back to Maxwell’s discussions about the tumbling of a thin card or sheet dropped in air \citep{maxwell1854particular}. Over recent decades, work on the so-called falling paper problem has greatly intensified due to interest in related forms of biological locomotion \citep{dickinson1999wing, walker2002rotational,bergou2010fruit,ristroph2011paddling, wang2016insect}. The aerodynamics of thin wings and plates undergoing unsteady motions has been interrogated by many methods, including direct numerical simulations via computational fluid dynamics \citep{sun2002unsteady, pesavento2004falling, wang2004unsteady, andersen2005unsteady}, laboratory experimentation \citep{birch2003influence, sane2001control,  andersen2005unsteady, dickson2004effect, huang2013experimental, li2022centre}, and mathematical modeling \citep{sane2002aerodynamic, birch2003influence, dickson2004effect, andersen2005analysis, andersen2005unsteady, pesavento2006unsteady,  hu2014motion, nakata2015cfd, li2022centre}. These and related studies have sought to characterize force generation mechanisms unique to the flight regime, such as the effect of leading edge vorticity and its shedding during wing translation as well as lift modifications due to pitching rotations \citep{dickinson1999wing, walker2002rotational, sane2002aerodynamic,   wang2004unsteady, fung2008introduction}. This line of work complements related research into the falling motions of sedimenting plates at lower $Re<10^2$ \citep{assemat2012onset, sun2024dynamics} as well as aeronautical research on plate-wings at high $Re>10^5$ \citep{tobak1981aerodynamic, tobak1985nonlinear, pinsky1994analysis, goman1997application, sinha2021elementary}.

A major goal has been to formulate mathematical force laws for the various contributing effects during flight and to incorporate these into dynamical models for the free motions of plate-wings \citep{farren1935reaction, sane2002aerodynamic, wang2012unsteady, wang2016insect}. These efforts parallel lift-drag types of laws and flight dynamics models of fixed-wing aircraft \citep{lee2016quasi, wang2012unsteady}. Given the intrinsic unsteadiness in the motions, flows, and forces during passive flight, the suitability of such a framework for the falling plate problem is not clear a priori. However, there have been notable successes with quasi-steady aerodynamic models that express forces in terms of instantaneous kinematic state variables, i.e. the plate’s orientation or attack angle, translational and rotational velocities, etc. \citep{andersen2005analysis, pesavento2006unsteady, wang2012unsteady, huang2013experimental, hu2014motion, nakata2015cfd}. Recent work by \citet{li2022centre} represents the current state of the art for models of the two-dimensional (2D) problem pertaining to planar motions of a thin plate, a setting that is recognized as involving much of the essential physics \citep{andersen2005unsteady,wang2016predictive, wang2016insect}. This nonlinear model built on and extended previous work to account for lift, drag, and added-mass effects associated with translation and rotation, as well as the torques associated with a dynamic center of pressure. The latter was shown to be important to account for the rich variety of motions displayed by plates of differing centers of mass, including end-over-end tumbling, back-and-forth fluttering, phugoid-like bounding, gliding and downward diving. Such states manifest differently in experiments with plastic plates falling in water and paper sheets in air, and the model was shown to successfully account for observations across these systems \citep{li2022centre}. 

As more observations become explainable by flight models, and more aerodynamic effects and conditions are subsumed within a single framework, new questions arise and new research directions become available. These include aspects of how the many different physical parameters defining the plate-fluid system map to the passive-flight states. Models are particularly well suited to address such issues given their efficiency, computational ease compared to direct numerical simulations, and versatility compared to experiments \citep{sane2002aerodynamic, dickson2004effect, andersen2005unsteady, wang2016insect}. Further, exploration of a given model's solution space should furnish new predictions that can be tested against other methods and therefore establish its applicability in different parameter regimes or drive its further refinement. Such work is motivated by the many potential applications. Quasi-steady modeling has already proven highly effective for insect flight \citep{birch2003influence, liu2005simulation, bergou2010fruit, ristroph2010discovering, ristroph2011paddling, ristroph2013active}, and other forms of motion and locomotion through air and water such as plant seed dispersal or finned propulsion may similarly benefit \citep{liu2005simulation,  miller2012using, wang2018three, certini2023flight}. For engineered systems such as small-scale flying and swimming vehicles and robots, accurate models could accelerate the design process and integrate into actuation and control schemes \citep{ellington1999novel,keennon2012development,ristroph2014stable,jafferis2019untethered}.  

In this work, we build on the model of \citet{li2022centre} to undertake an exploration of the space of passive-flight patterns across the widely ranging scales and conditions commonly accessed in aerial and aquatic environments. Dimensional analysis allows us to reduce the complexity of the parameter space for plates of various physical characteristics moving through fluids of differing material properties, and stability analysis of equilibrium solutions to the model yields maps that help to predict and characterize the flight behaviors. These investigations show that the full gamut of flight motions arise across the space of parameters, and that any given state such as gliding can be achieved in distinct ways. This work also spurs useful refinements of the model, furnishes formulas for the stability of steady motions, and leads to predictions of new unsteady motions, whose existence may be validated or refuted in future experiments and/or direct numerical simulations. Overall, these results reveal an unexpectedly complex space of passive-flight behaviors that can, however, be organized and understood through the presented modeling and analysis techniques.

\section{Dimensional and scaling analyses}

\begin{figure}
\centering
    \includegraphics[scale=0.5]{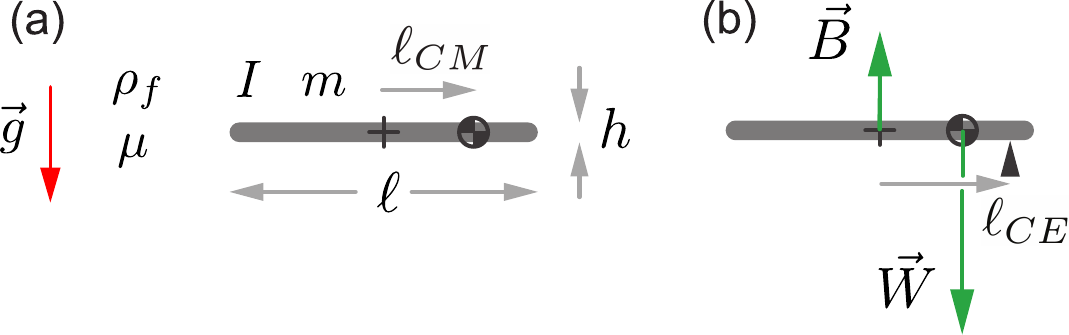}
  \caption{Quantities relevant to the passive flight of a thin plate. (a) A plate of length $\ell$ and thickness $h$ has mass $m$, center of mass $\ell_{CM}$ measured from the middle, and moment of inertia $I$. It moves under the action of gravity (acceleration $\vec{g}$) through an ambient fluid of density $\rho_f$ and viscosity $\mu$. (b) The center of static equilibrium $\ell_{CE}$ is defined as the balance point for the torques due to weight and buoyancy.}
  \label{fig:plate_quantities}
\end{figure}

We seek to establish dimensionless groups of variables with which to describe the general problem of a rigid plate of arbitrary mass distribution that passively falls under gravity through a Newtonian fluid. We follow previous works and address the two-dimensional problem \citep{andersen2005analysis, li2022centre, sane2002aerodynamic, hu2014motion}. The situation is characterized by the 8 dimensional quantities shown in figure \ref{fig:plate_quantities}a. There are 5 quantities intrinsic to the plate: chord length, $\ell$; thickness, $h$; center of mass, $\ell_{CM}$; the 2D mass, $m$, as measured per unit span; and 2D moment of inertia, $I$. There are 3 additional environmental quantities: fluid density, $\rho_f$; dynamic viscosity, $\mu$; and the gravitational acceleration, $\vec{g}$. These 8 total quantities may be reduced by the Buckingham $\pi$ theorem to 5 dimensionless groups \citep{logan2013applied}. Reasonable choices consistent with the aerodynamics literature are \citep{andersen2005analysis, li2022centre}:
\begin{equation} \label{dim_all1}
    \ell_{CM}^* = \frac{\ell_{CM}}{\ell}, \quad h^* = \frac{h}{\ell}, \quad M^* = \frac{m}{\pi\rho_f \left(\frac{\ell}{2}\right)^2}, \quad I^* = \frac{I}{\frac{1}{2}\pi\rho_f \left(\frac{\ell}{2}\right)^4}, \quad \Rey = \frac{\sqrt{2mg\rho_f\ell}}{\mu}.
\end{equation}
Respectively, these correspond to the normalized center of mass, the thickness aspect ratio, the mass of the plate relative to the fluid, the relative moment of inertia, and the Reynolds number $\Rey=\rho_f U \ell / \mu$ based on a speed scale $U = \sqrt{2mg/\rho_f\ell}$ set by balancing weight with a fluid force of the usual high-$Re$ form that increases quadratically with speed.

Anticipating applications for different fluids, we consider a related set of parameters that explicitly includes the effect of buoyancy. To this end, it is convenient to define the dimensionless form of the buoyancy-corrected weight $W^* = (W-B)/W = 1-\rho_{f}hl/m = 1-4h^*/\pi M^*$. The last two expressions hold specifically for a plate of rectangular cross-section, and the final form indicates that $W^*$ may replace $h^*$ in the dimensionless set of variables. Further, the Reynolds number is readily modified by replacing the weight with its buoyancy-corrected form: $mg \rightarrow W^* m g$ \citep{amin2019role}. We also replace the center of mass with the more general center of static equilibrium, which is the location at which the torques due to buoyancy and weight balance in a static situation without flow \citep{li2022centre}. Torque balance based on the force diagram of figure \ref{fig:plate_quantities}b leads to the relation $\ell_{CE} = \ell_{CM}/W^*$. In summary, the selected 5 dimensionless groups can be expressed in terms of the 8 dimensionful plate-fluid quantities as:
\begin{equation}\label{dim_all2}
    \ell_{CE}^* = \frac{\ell_{CM}}{W^* \ell},~~ W^* = 1-\frac{\rho_{f}hl}{m},~~ M^* = \frac{m}{\pi\rho_f \left(\frac{\ell}{2}\right)^2},~~ I^* = \frac{I}{\frac{1}{2}\pi\rho_f \left(\frac{\ell}{2}\right)^4},~~ \Rey = \frac{\sqrt{2W^* mg\rho_f\ell}}{\mu}.
\end{equation}
These parameter definitions are summarized in table \ref{dimensionless} along with their ranges.

The variables $\ell_{CE}^*$, $W^*$, $M^*$, and $I^*$ will appear throughout our study, and we will examine their effect within a model. The parameter $\Rey$ will not appear explicitly since the aerodynamic coefficients (e.g., lift, drag, and added mass) are assumed to be independent of Reynolds number over the intermediate range ($10^2$ to $10^5$) of interest here. The model and our results strictly pertain to thin plates in accordance with the expressions in equation \ref{dim_all2} and with the assumption that the aerodynamic coefficients are independent of the slenderness ratio $h/\ell$. The work of \citet{li2022centre} showed good agreement of such a model with experiments conducted for $h/\ell = 0.001$ and $0.1$. The form of the model may be more generally relevant to the flight of slender wings whose shape and Reynolds number would dictate the model coefficients. One may also reasonably apply our results to flight systems composed of a thin wing as the aerodynamically relevant surface and other structures that experience significantly weaker fluid forces but nonetheless contribute (perhaps strongly) to mass, buoyancy, center of mass, moment of inertia, etc. For example, a gliding bird could be crudely viewed in this way as composed of wings and a body (fuselage).

\begin{table}
\centering
\begin{tabular}{|c c c|} 
 Quantity & Definition & Range \\
 \hline
 Center of equilibrium, $\ell_{CE}^*$ & $\frac{\ell_{CM}^*}{W^*}$ & $[0,\infty)$ \\
  Effective weight, $W^*$ & $1-\frac{\rho_f h\ell}{m}$ & $(0,1)$\\
  Mass, $M^*$ & $\frac{m}{\pi\rho_f(\ell/2)^2}$ & $(0,\infty)$ \\
    Moment of inertia, $I^*$ & $\frac{2I}{\pi\rho_f(\ell/2)^4}$ & $(0,\infty)$ \\
 Reynolds number, $\Rey$ & $\frac{\sqrt{2W^*mg\rho_f\ell}}{\mu}$ & $(10^2,10^5)$ 
\end{tabular}
\caption{Summary of dimensionless quantities and their ranges for the problem of a thin plate falling passively under gravity through fluid.}
\label{dimensionless}
\end{table}

\begin{figure}
\centering
    \includegraphics[scale=0.5]{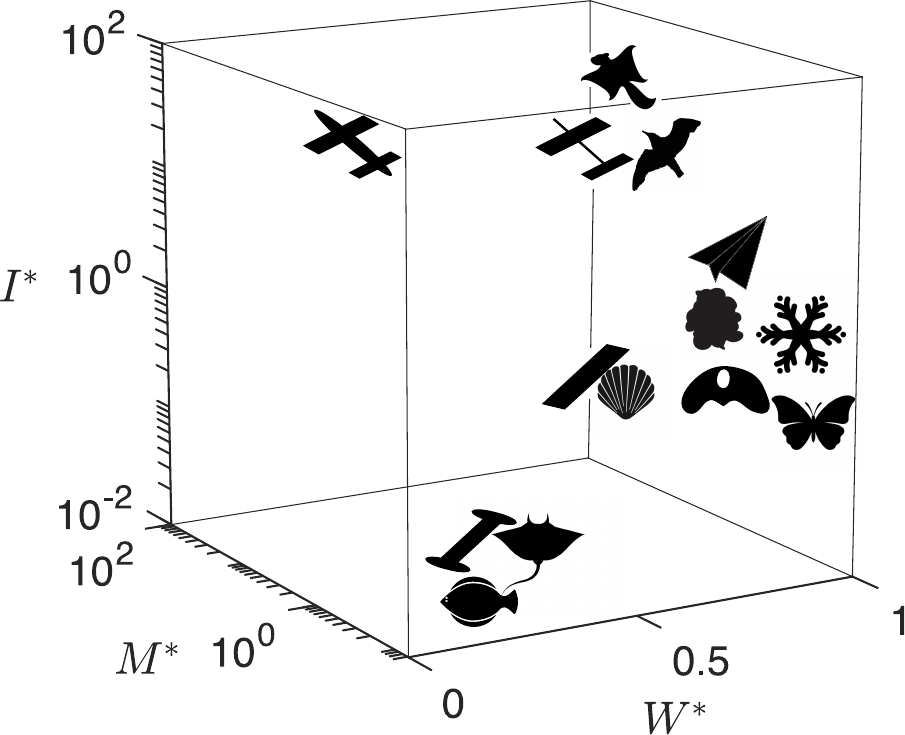}
  \caption{Real-world flight systems occupy different regions of the parameter space. The selected examples vary in size and composition and occupy either air or water, and they rely on thin structures operating at intermediate Reynolds numbers. Proceeding roughly from top to bottom, shown are a flying squirrel, an autonomous gliding water craft, a gliding air vehicle, a bird, a paper airplane, a particulate of marine snow, a snowflake, an aluminum plate in water, a scallop, a seed, a butterfly, an acrylic plate in water, a stingray in water, and a fish.}
  \label{fig:system}
\end{figure}

\subsection{Survey of intermediate-$Re$ passive fliers}

The need for a general analysis of the passive flight problem is motivated by the widely ranging parameter values characterizing the relevant systems. In figure \ref{fig:system} we place some representative fliers on the 3D map defined by the quantities $(W^*,M^*,I^*)$, whose values can be estimated from information in the literature (Appendix A). The examples displayed include laboratory idealizations involving metal or plastic plates in water; the everyday case of paper in air; flying animals such as insects, mammals and birds; swimming animals such as mollusks, fish and rays; plant seeds; biomimetic robots; and abiotic fliers such as marine snow and airborne snowflakes. What these have in common is that all have thin structures that dictate their intermediate-$Re$ motions through fluids. Note that the parameter $\ell_{CE}^*$ is not shown since reliable data are generally not available. In addition, the displayed parameters should be understood as rough estimates, and hence the figure is intended only to convey that the relevant values span orders of magnitude.

Some further details give greater appreciation for the diversity among the relevant systems. Proceeding generally top to bottom, shown in figure \ref{fig:system} are a flying squirrel (Glaucomys volans) \citep{thorington1981body}, an autonomous gliding water vehicle \citep{wood2009autonomous}, an autonomous gliding air vehicle \citep{wood2007autonomous}, a bird (Uria aalge) \citep{berg1995moment}, a paper airplane \citep{li2022centre}, a flake of marine snow \citep{passow2012marine}, a snowflake \citep{langleben1954terminal}, an aluminum plate in water \citep{andersen2005unsteady}, a scallop (Placopecten magellanicus) \citep{cheng1996dynamics}, a seed (Alsomitra macrocarpa) \citep{viola2022flying, ennos1989effect}, a butterfly (Papilio ulysses) \citep{hu2010experimental}, a plastic plate in water \citep{li2022centre}, a stingray in water (Dasyatis pastinaca) \citep{yigin2012age}, and a flounder (Paralichthys olivaceus) \citep{takagi2010functional}. As detailed in the Appendix, the parameters $(W^*,M^*,I^*)$ are computed as order-of-magnitude estimates based on reported values of the relevant morphological parameters and dimensions.

Can a single model usefully apply across such large variations in parameter values? Are steady motions such as gliding only available in restricted regions of the space, or are they generally accessible? How does the stability of such states vary with the parameters? These are some of the questions motivating the work presented here.

\section{Existence and uniqueness of free-flight equilibrium states}

\begin{figure}
\centering
    \includegraphics[scale=0.48]{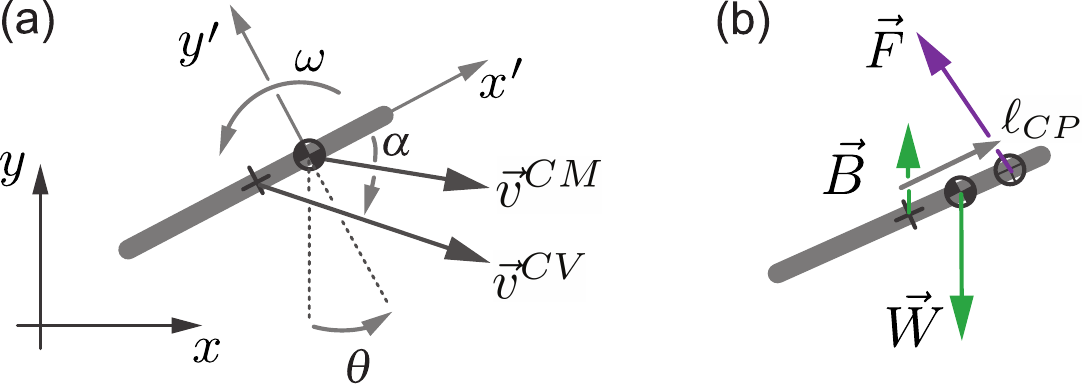}
  \caption{Definitions of dynamic quantities. (a) A snapshot of the plate. The center of mass has location $(x,y)$ in the lab frame and translates with velocity $\vec{v}^{CM}$, and the plate has posture given by the angle $\theta$ and rotates with angular velocity $\omega$. The primed frame co-rotates with the plate, and the angle of attack $\alpha$ is that of the center of volume velocity $\vec{v}^{CV}$ relative to the $x'$ axis. (b) Free body diagram of the forces. The net aerodynamic force $\vec{F}$ acts at the center of pressure $\ell_{CP}$, the weight $\vec{W}$ acts at the center of mass, and the buoyancy $\vec{B}$ acts at the center of volume (middle).}
  \label{fig:plate_quantities_dynamic}
\end{figure}

Our characterizations start by seeking to identify the equilibrium flight states that satisfy zero net force and torque and which therefore involve steady translation and rotation. Such motions are specified by the constant values of 3 kinematic variables, which would generally be the rotation rate $\omega$ and the two components of the center of mass velocity $\vec{v}^{CM}=(v^{CM}_{x},v^{CM}_{y})$. These and other free-flight variables are defined in figure \ref{fig:plate_quantities_dynamic}a. However, rotation can be immediately excluded on the basis that such states generally involve time-dependent fluid forces and torques. It is therefore the special cases with $\omega=0$ that are of interest, for which we instead choose the orientation angle $\theta$ defined relative to the vertical and the velocity $\vec{v}$, which is the same for any point on the wing. Equivalently, the state is specified by $\theta$, the speed $v = |\vec{v}|$, and angle of attack $\alpha$ from the plate surface to its velocity vector. These 3 kinematic constants combine with the many dimensional parameters (8 for rectangular plates) characterizing the wing-fluid system by a grand total of 11 quantities which specify the wing-fluid-flight system. These quantities are not independent, and we seek a minimal subset of parameters whose values must be specified in order to determine the others.

Here we claim a correspondence principle that we will later show holds mathematically within our model and which may apply in a looser sense to real flight systems: \textit{specifying the static quantities that arise for a wing fixed in a wind tunnel setting determines the full set of dynamical quantities involved in its free-flight equilibrium at the same relative flow conditions.} That is, for a wing of given geometry (shape and size) fixed in a given fluid (density and viscosity) under steady flow conditions (attack angle and speed), one can determine the remaining factors (orientation and trajectory of the wing, its mass, center of mass, and moment of inertia) needed to achieve the same aerodynamic conditions as an equilibrium motion through the fluid under gravity. This equilibrium may or may not be aerodynamically stable and hence may or may not be maintained as a free flight state.

\begin{figure}
\centering
    \includegraphics[scale=0.6]{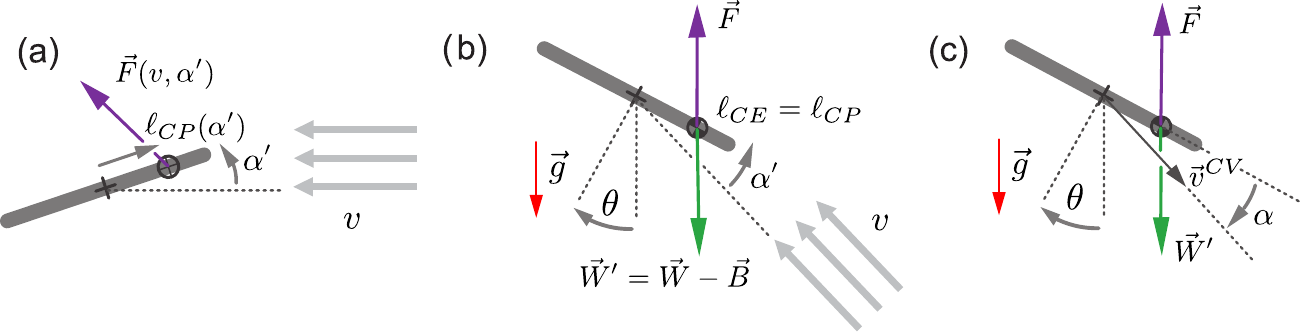}
  \caption{The aerodynamic conditions of a fixed wing in a wind tunnel flow can be identically realized as an equilibrium state of free flight. (a) A plate held fixed at attack angle $\alpha'>0$ in a wind tunnel of flow speed $v$ experiences an aerodynamic force $\vec{F}(v,\alpha')$ that acts at the center of pressure $\ell_{CP}(\alpha')$. (b) The entire plate-tunnel system can be rotated so that $\vec{F}$ points directly upward, which determines the orientation angle $\theta$. The plate can then be released under gravity, and its total mass may be assigned such that $|\vec{W}'| = |\vec{F}|$ in order to achieve force balance. The mass may be distributed such that $\ell_{CE}=\ell_{CP}$, which ensures torque balance. (c) A change of reference frames indicates that the same conditions can be achieved as an equilibrium motion through quiescent fluid. Note the wind-tunnel convention defines the attack angle $\alpha'$ as that of the plate relative to the upstream direction, whereas the free-flight angle $\alpha=-\alpha'$ is defined here as that of the plate velocity vector relative to the $x'$ axis. }
    \label{fig:proof}
\end{figure}

The argument starts in the wind tunnel setting, where a wing is held fixed in a steady uniform flow of speed $v$ at angle of attack $\alpha'$, defined as that of the plate relative to the flow as shown in figure \ref{fig:proof}a. These factors determine the aerodynamic force $\vec{F}$ due to pressure and the center of pressure $\ell_{CP}$, which is the effective point of action. The entire wing-tunnel system can be rotated to make $\vec{F}$ point vertically upward as in figure \ref{fig:proof}b. This determines the plate orientation angle $\theta$ while keeping $\alpha'$ and $v$ at their prescribed values. The wing may then be released from rest within this inclined flow. Force balance is achieved only if the fluid force balances the buoyancy-corrected weight, $F = W-B$, which therefore determines the mass $m$ since the buoyancy is fixed by the prescribed geometry. Torque balance is achieved only by matching the pressure and equilibrium centers, $\ell_{CP} = \ell_{CE}$, which determines the center of mass $\ell_{CM}$. The wing may now hover in place within the inclined flow. Invoking Galilean relativity as in figure \ref{fig:proof}c, the same aerodynamic conditions are realized in free flight through a quiescent fluid by a downward trajectory of the wing at speed $v^{CV}=v$ and $\alpha=-\alpha'$, defined here as that of the velocity vector relative to the plate. The moment of inertia $I$ is undetermined but irrelevant, i.e., any identified equilibrium may be achieved for any value of $I>0$. This specifies the sense in which free-flight equilibria exist and are unique.

Note that $\alpha'=0$ is a special case in that there is no torque, $\ell_{CP}$ is undefined, and $\ell_{CE}$ is thus undetermined. While any location of the center of mass is permissible, the other free-flight factors such as mass, velocity, and attack angle are determined according to the reasoning given above.

In summary, this argument takes as inputs the wing geometry (for a plate, $\ell$ and $h$), the fluid and environmental parameters ($\rho_f$, $\mu$ and $g$), and the usual wind tunnel quantities ($\alpha'$ and $v$) and from these determines the remaining quantities ($\theta$, $m$ and $\ell_{CM}$, with $I$ free) needed to completely specify a free-flight equilibrium state. This argument is not unique, e.g., one could take the plate mass $m$ and free-flight attack angle $\alpha$ as inputs to determine the speed $v$. We will later analyze equilibria of a dimensionless version of a flight model to show that $\alpha$ can be taken as the sole input that determines $(\theta,\vec{v}^{CV*},\ell_{CE}^*)$, with $W^*$, $I^*$ and $M^*$ being free. The resulting map $(\alpha, W^*, M^*, I^*)$ to $(\theta, \vec{v}^{CV*},\ell_{CE}^*)$ is then exploited to simplify the stability analysis of the equilibria.

Implicit in the above argument are conditions that are often assumed in aerodynamic contexts and which will be exactly satisfied within our quasi-steady framework. The fluid force and torque must derive dominantly from pressure, as expected for sufficiently high $\Rey>10^2$ \citep{tritton2012physical}. The forces are assumed steady, as expected for low $\alpha$ and stably attached boundary layer flows at sufficiently low $\Rey<10^5$ \citep{schlichting2016boundary,anderson2011ebook}. (Alternatively, one may consider the force balances defining the equilibria as applying in the time average.) Slenderness of the wing is needed so that the center of pressure is well defined via integrals along the centerline of the pressure difference across the surface: $\ell_{CP} = \int xp(x)dx / \int p(x)dx$. For a given wing geometry, the pressure center $\ell_{CP}(\alpha')$ is assumed to depend only on the attack angle. Finally, for a given wing at fixed $\alpha'$ in a given directed flow, the fluid force $\vec{F}$ is assumed to have a unique direction, and its magnitude has a one-to-one (bijective) relationship with the relative flow speed. Such conditions are expected insofar as the flow state is unique for a given set of conditions and for pressure forces that typically increase quadratically with speed at moderate to high $\Rey$  \citep{tritton2012physical,anderson2011ebook}.

\subsection{Equilibrium states for plates}

\begin{figure}
\centering
    \includegraphics[scale=0.4]{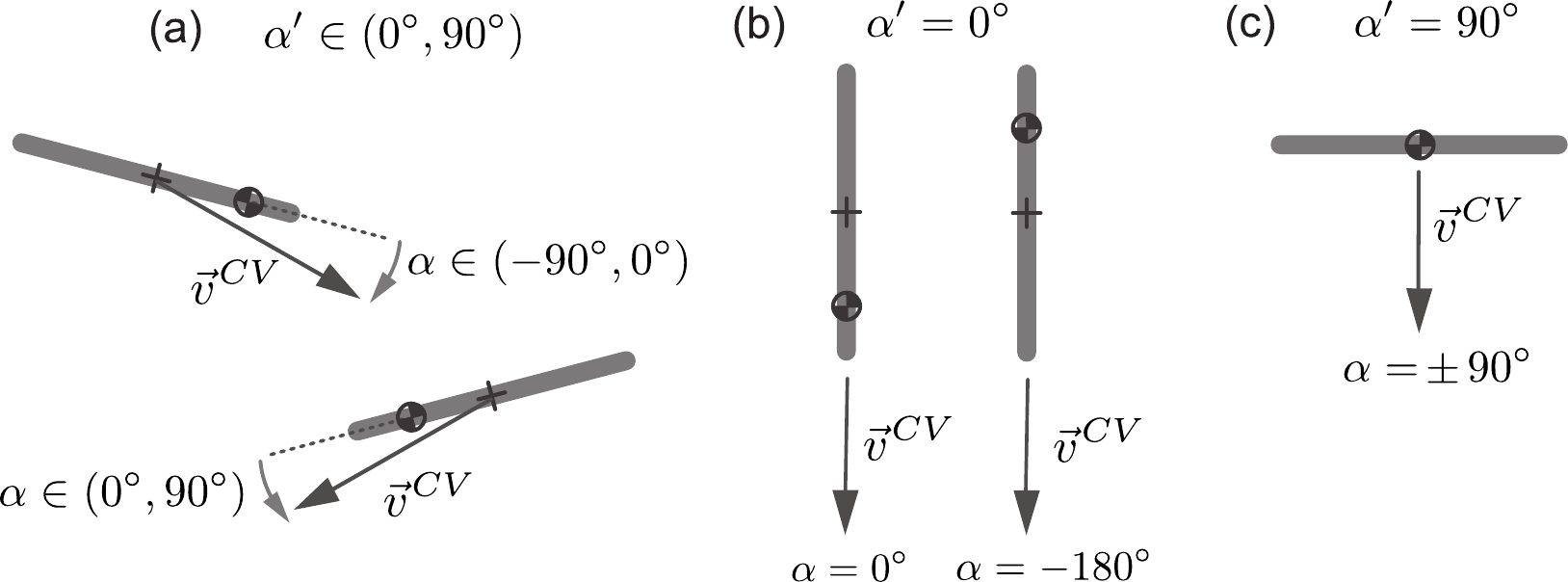}
  \caption{Three types of free-flight equilibria of a plate. (a) Gliding involves constant speed motion along a sloped trajectory and with an acute angle of the plate. Each attack angle $\alpha'\in(0,\pi/2)=(0^\circ,90^\circ)$ as conventionally defined for the wind tunnel setting admits two free-flight states with $\alpha=\pm\alpha'$ corresponding to leftward and rightward gliding. (b) Diving involves constant speed descent directly downward and with edgewise posture of the plate. For a given value of $\ell_{CE}^*\geq 0$, two such states exist. (c) Pancaking involves constant speed descent directly downward and with broadside-on posture. This is achieved only for $\ell_{CE}^*=0$, and thus the two states are physically identical and degenerate. }
  \label{fig:gliding_diving}
\end{figure}

The fixed- and free-flight correspondence principle readily allows all equilibria to be tabulated based on the attack angle, and the possibilities are depicted in figure \ref{fig:gliding_diving}. For this purpose, it is helpful to distinguish two notions of the angle of attack. The static angle $\alpha' \in [0,\pi/2] =[0^{\circ},90^{\circ}]$ is relevant to the fixed configuration of a wind tunnel, where the limited range is sufficient for completely specifying the aerodynamic properties of a plate given its fore-aft and up-down symmetries. The dynamic angle $\alpha \in [-\pi,\pi) = [-180^{\circ},180^{\circ})$ is relevant to free-flight, where the full range is needed to completely specify the flight state and allow for all possible directions of the velocity vector relative to the plate's $x'$-axis.

We use the term \textit{gliding} for those states with strictly acute static angle $\alpha' \in (0,\pi/2) =(0^{\circ},90^{\circ})$, as shown in figure $\ref{fig:gliding_diving}$a. For a given static $\alpha'$, there are two available free-flight motions that take the form of leftward and rightward descent along linear trajectories for which the dynamic angles are $\alpha = \pm \alpha'$. It will be shown that gliding at a given $\alpha'$ is associated with a unique value of the equilibrium center $\ell_{CE}^*$. 

We introduce special terms for the equilibria at the two extremes of $\alpha'$. We use the term \textit{diving} for those states with $\alpha'= 0 = 0^{\circ}$, which involve edge-on and strictly downward descent as shown in figure \ref{fig:gliding_diving}b. For a given $\ell_{CE}^*>0$, there are two available free-flight motions that take the form of bottom-heavy diving with $\alpha= 0 = 0^{\circ}$ and top-heavy diving with $\alpha= -\pi = -180^\circ$. The two are distinguished by whether the center of equilibrium is displaced towards the leading or trailing edge, and they are degenerate for the symmetric case of $\ell_{CE}^*=0$. Diving is exceptional in that torque balance is achieved for all values of $\ell_{CE}^*\geq0$, as the weight and aerodynamic force both act parallel to the plate, and it will necessitate a separate analysis of stability. We use the term \textit{pancaking} for those states with $\alpha'=\pi/2=90^{\circ}$, which involve broadside-on and strictly downward descent as shown in figure \ref{fig:gliding_diving}c. These states will be shown to exist only for the symmetric case $\ell_{CE}^*=0$. In principle, they take the form of two motions with $\alpha=\pm\pi/2=\pm90^\circ$, which however are degenerate and physically indistinguishable. Within our quasi-steady model and its analysis, pancaking is simply a particular case of gliding that requires no special treatment.

\section{Flight dynamics model and numerical solutions}

We propose and analyze a dynamical system for the problem of a falling plate that builds on and extends the work of \citet{li2022centre}. The model expresses the Newton-Euler equations for planar (2D) motion with forces and torques due to gravity (weight), fluid-static effects (buoyancy) and fluid-dynamic effects (pressure, skin friction, added mass, etc.). As shown in figure \ref{fig:plate_quantities_dynamic}a, the plate has center-of-mass position $(x,y)$ in the lab (fixed) frame and center-of-mass velocity $\vec{v}^{CM} = (v_{x}^{CM},v_{y}^{CM})$. Its instantaneous orientation angle is $\theta$ and its angular velocity is $\omega$. It proves most convenient to express the dynamical variables in a frame that rotates with the plate, e.g., $(v_{x'}^{CM},v_{y'}^{CM})$, where the prime indicates the co-rotating frame (figure \ref{fig:plate_quantities_dynamic}a). The aerodynamic forces and torques are expressible in terms of the motion of the geometric middle or center of volume CV,
\begin{equation}
v_{x'}^{CV} = v_{x'}^{CM} = v_{x'} \quad \textrm{and} \quad v_{y'}^{CV} = v_{y'}^{CM}-\omega\ell_{CM} = v_{y'}-\omega\ell_{CM},
\end{equation}
where we suppress the superscript CM hereafter for ease of notation. The dynamics take the form of a system of nonlinear, coupled ordinary differential equations (ODEs) whose dimensional form is given by:
\begin{equation}\label{ODEsystem_original}
\begin{split}
\dot{x} &= v_{x'}\cos\theta-v_{y'}\sin\theta\\
\dot{y} &= v_{x'}\sin\theta+v_{y'}\cos\theta\\
\dot{\theta} &= \omega\\
(m+m_{11})\dot{v_{x'}}&=\left(m+m_{22}\right)\omega v_{y'}-m_{22}\omega^2\ell_{CM}+L_{x'}+D_{x'}-m'g\sin\theta\\
\left(m+m_{22}\right)\dot{v_{y'}} &= -(m+m_{11})\omega v_{x'} + m_{22}\dot{\omega}\ell_{CM} + L_{y'} + D_{y'} - m'g\cos\theta\\
\left(I + I_a\right)\dot{\omega} &= \tau_T + \tau_{RL} + \tau_{RD} +\tau_B.
\end{split}
\end{equation}
This model is identical to that of \citet{li2022centre} with the exception of the term $\tau_{RL}$, which is newly added here and will be discussed below. The first three equations relate positions and angle to their respective velocities, and the last three equations are the Newton-Euler equations for the accelerations induced by forces and torques. Added mass effects are associated with terms involving the coefficients $m_{11} = 0$, $m_{22} = \pi\rho_f\ell^2/4$ and $I_{a} = I_a(\ell_{CM}=0) + m_{22}\ell_{CM}^2 = \pi\rho_f\ell^4[1+8(2\ell_{CM}/\ell)^2]/128$, where the expressions hold for an infinitesimally thin plate. The lift $\vec{L}$ and drag $\vec{D}$ terms are detailed below and expressed in terms of the velocities and posture of the plate, with force coefficients that were empirically determined by \citet{li2022centre} for intermediate $\Rey$. Aerodynamic effects also induce torques that are decomposed into $\tau_T$, $\tau_{RL}$ and $\tau_{RD}$ according to their association with lift and drag from wing translation (T), lift from rotation (RL) and drag from rotation (RD). Finally, buoyancy effects are accounted for in the corrected mass $m' = (\rho_s-\rho_f)V = (\rho_s-\rho_f)h\ell$ for a plate of homogeneous density $\rho_s$ and 2D volume $V$, as well as in the torque $\tau_B$ about the center of mass. These quantities are defined as follows:

\begin{equation}\label{dimensional_ODE_quantities}
\begin{split}
\vec{L}_T &=  \frac{1}{2}\rho_f\ell C_L(\alpha)\sqrt{v_{x'}^2 + (v_{y'}-\omega\ell_{CM})^2}(v_{y'}-\omega\ell_{CM}, -v_{x'})\\
\vec{L}_R &=  -\frac{1}{2}\rho_f\ell^2C_R\omega(v_{y'}-\omega\ell_{CM}, -v_{x'})\\
\vec{L} &= \vec{L}_T + \vec{L}_R \\
\vec{D} &= -\frac{1}{2}\rho_f\ell C_D(\alpha)\sqrt{v_{x'}^2 + (v_{y'}-\omega\ell_{CM})^2}(v_{x'}, v_{y'}-\omega\ell_{CM})\\
\tau_T &= -\frac{1}{2}\rho_f\ell \sqrt{v_{x'}^2+(v_{y'}-\omega\ell_{CM})^2}\left[C_L(\alpha)v_{x'}+C_D(\alpha)(v_{y'}-\omega\ell_{CM})\right]\left[\ell_{CP}(\alpha)-\ell_{CM}\right]\\
\tau_{RL} &= \frac{1}{2}\rho_f\ell C_R\omega v_{x'}(\ell_{CM}-\ell_{CRL})\\
\tau_{RD} &=  -\frac{1}{128}C_D^{\pi/2}\omega|\omega|\left[(2\ell_{CM}/\ell+1)^4 + (2\ell_{CM}/\ell-1)^4\right]\\
\tau_B &= - \rho_f gh\ell\ell_{CM}\cos\theta.
\end{split}
\end{equation}

\begin{figure}
\centering
    \includegraphics[scale=0.55]{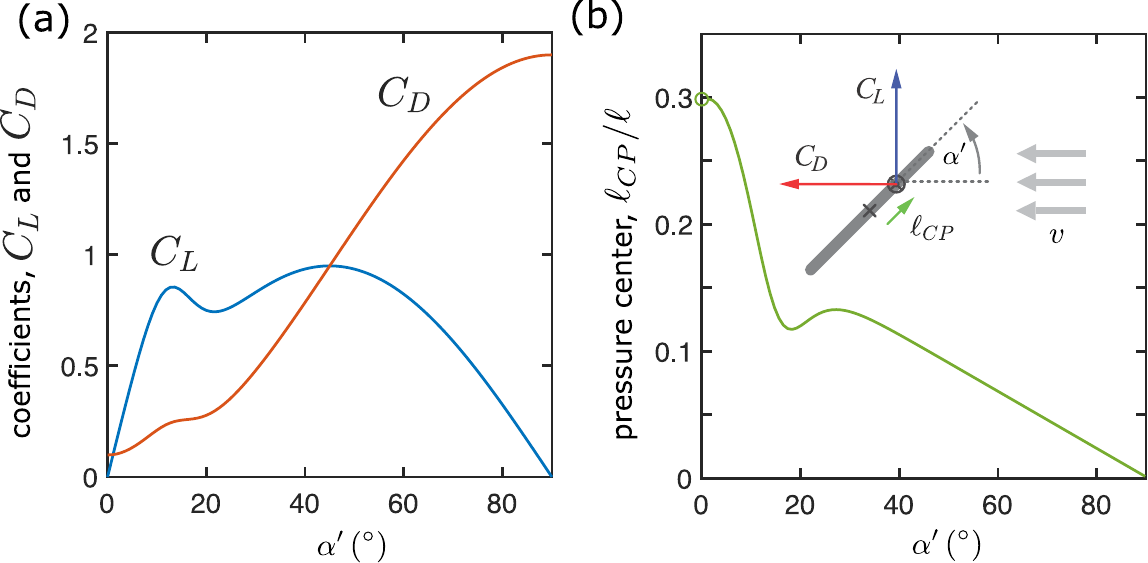}
  \caption{Aerodynamic force characteristics of a thin plate at intermediate $Re$ as determined by the experimental tunnel measurements of \citet{li2022centre}. (a) Lift and drag coefficients as functions of the attack angle $\alpha' \in [0,\pi/2]=[0^\circ,90^\circ]$, whose range covers all unique postures relative to the flow. Stall is evident in the drop in lift near $\alpha' = 15^\circ$. (b) The center of pressure location along the plate. Stall leads to a non-monotonic form of the curve. The value at $\alpha=0$ is undefined as the force is parallel to the plate surface.}
  \label{fig:lift_drag}
\end{figure}

The various aerodynamic coefficients are largely taken from \citet{li2022centre}, where they were determined by theoretical considerations and experimental measurements. The rotational lift coefficient $C_R = 1.1$ is taken is a constant, which was shown in previous models to adequately reproduce observations from experiments and direct numerical simulations \citep{andersen2005analysis, andersen2005unsteady, pesavento2004falling, pesavento2006unsteady}. Lacking any information on the center of rotational lift, we take it to be $\ell_{CRL} = 0$. Other quantities are assumed to depend on the dynamic attack angle $\alpha = \arctan[(v_{y'}-\ell_{CM}\omega)/v_{x'}]$, including the lift coefficient $C_{L}(\alpha)$, drag coefficient $C_{D}(\alpha)$ and center of pressure $\ell_{CP}(\alpha)$. The following expressions from \citet{li2022centre} are appropriate for angles $\alpha \in [0,\pi/2] = [0^\circ,90^\circ]$: 
\begin{equation}
    \begin{split}
        C_L(\alpha) &= f(\alpha)C_L^1\sin\alpha + [1-f(\alpha)]C_L^2\sin(2\alpha) \\
        C_D(\alpha) &= f(\alpha)(C_D^0 + C_D^1\sin^2\alpha) + [1-f(\alpha)]C_D^{\pi/2}\sin^2\alpha \\
        \ell_{CP}(\alpha)/\ell &= f(\alpha)(C_{CP}^0-C_{CP}^1\alpha^2)+[1-f(\alpha)]C_{CP}^2[1-\alpha/(\pi/2)]\\
        f(\alpha) &= [1-\tanh(\alpha-\alpha_0)/\delta]/2.
    \end{split}
\end{equation}
Here, the constant prefactors are $C_L^1 = 5.2$, $C_L^2 = 0.95$, $C_D^0 = 0.1$, $C_D^1 = 5.0$, $C_D^{\pi/2}=1.9$, $C_{CP}^{0}=0.3$, $C_{CP}^{1}=3.5$, $C_{CP}^{2}=0.2$, $\alpha_0=14^{\circ}$ and $\delta=6^{\circ}$. The logistic function $f(\alpha)$ plays the mathematical role of an indicator function that smoothly transitions between different expressions appropriate for low $\alpha < \alpha_0$ where one expects an attached leading-edge vortex or so-called separation bubble and higher $\alpha > \alpha_0$ where the flow fully separates and stall occurs \citep{smith2021scales}. The aerodynamics therefore differs markedly from classical airfoil theory \citep{anderson2005introduction, anderson2011ebook}. Figure \ref{fig:lift_drag} shows the corresponding curves identified in water tunnel experiments by \citet{li2022centre}, where they were shown to account for experimental observations on plates of thickness ratios $h/\ell = 10^{-3}$ to $10^{-1}$ and Reynolds numbers $Re = 10^2$ to $10^4$. The above expressions are readily extended throughout $\alpha\in[-\pi,\pi) = [-180^\circ,180^\circ)$ based on symmetries as explained in \citet{li2022centre}, which allows the model to address arbitrary motions during free flight.

\subsection{Torque from rotational lift}

We give extra consideration to the torque from rotational lift $\tau_{RL}$ as it is the only new addition to the model presented in \citet{li2022centre}. That work included rotational lift $\vec{L}_R$ as a Magnus-like force that is associated with the combined translation and rotation of a wing and which scales with the product of the two respective speeds \citep{munk1925note,kramer1932zunahme,sane2003aerodynamics}. However, no associated torque was included, and indeed to our knowledge no previous work has addressed a possible torque contribution from this effect. Its omission in the model of \citet{li2022centre} is conceptually problematic, since the absence of an associated torque implies that this force always acts at the center of mass. This violates the fundamental physical principle that all fluid dynamical effects depend only on the outer shape and motion of a structure and do not directly ``know'' about aspects of mass and its distribution inside the structure.

We propose a remedy in which the rotational lift force $\vec{L}_R$ is associated with an effective point of action or center $\ell_{CRL}$. This is analogous to how pure translation gives rise to pressure forces (translational lift and drag) that act at the center of pressure. As such, the expression for $\tau_{RL} = L_{R_{y'}}(\ell_{CM}-\ell_{CRL})$ in equation \ref{dimensional_ODE_quantities} follows directly from that for $\vec{L}_R$. The center of rotational lift $\ell_{CRL}$ could in principle vary with attack angle and perhaps other dynamical quantities, but such information seems unavailable in the literature. To avoid introducing any unsubstantiated dependencies, we opt for the simplest choice of $\ell_{CRL}=0$, i.e., the rotational lift always acts at the middle of the plate. This choice could be viewed as consistent with previous models of \citet{andersen2005analysis,andersen2005unsteady} that included rotational lift without any associated torque for symmetrically weighted plates, which have $\ell_{CM}=0$ and hence $\tau_{RL}=0$ only if $\ell_{CRL}=0$. Future experiments or numerical simulations may provide more information about this effect, and the model may be updated accordingly.

\subsection{Dimensionless form of dynamical system}

Nondimensionalization of the ODE system leads to equivalent expressions that involve the aforementioned dimensionless variables $(\ell_{CE}^*,W^*,M^*,I^*,\Rey)$. We choose the characteristic length scale to be $\ell$ and time scale to be $\ell/U$, recalling the characteristic speed $U = \sqrt{2W^*mg/\rho_f\ell}$. The dynamical system then becomes:
\begin{equation}\label{ODEsystem_final}
\begin{split}
\dot{x}^* &= v_{x'}^*\cos\theta-v_{y'}^*\sin\theta\\
\dot{y}^* &= v_{x'}^*\sin\theta+v_{y'}^*\cos\theta\\
\dot{\theta} &= \omega^*\\
M^*\dot{v_{x'}}^*&=\left(1+M^*\right)\omega^* v_{y'}^*-(\omega^*)^2W^* \ell_{CE}^*+L_{x'}^*+D_{x'}^*-\frac{2}{\pi}\sin\theta\\
\left(1+M^*\right)\dot{v_{y'}}^* &= -M^*\omega^* v_{x'}^* + \dot{\omega}^*W^* \ell_{CE}^* + L_{y'}^* + D_{y'}^* - \frac{2}{\pi}\cos\theta\\
\left[I^* + \frac{1+32(W^* \ell_{CE}^*)^2}{4}\right]\dot{\omega}^* &= \tau_T^* + \tau_{RL}^* + \tau_{RD}^* +\tau_B^*.
\end{split}
\end{equation}
The aerodynamic forces and torques are given by:
\begin{equation}\label{lift_drag_torque}
\begin{split}
\vec{L}_T^* &=  \frac{2}{\pi}C_L(\alpha)\sqrt{(v_{x'}^*)^2 + (v_{y'}^*-\omega^*\ell_{CM}^*)^2}(v_{y'}^*-\omega^*\ell_{CM}^*, -v_{x'}^*)\\
\vec{L}_R^* &=  -\frac{2}{\pi}C_R\omega^*(v_{y'}^*-\omega^*\ell_{CM}^*, -v_{x'}^*)\\
\vec{L}^* &= \vec{L}_T^* + \vec{L}_R^* \\
\vec{D}^* &= -\frac{2}{\pi}C_D(\alpha)\sqrt{(v_{x'}^*)^2 + (v_{y'}^*-\omega^*\ell_{CM}^*)^2}(v_{x'}^*, v_{y'}^*-\omega^*\ell_{CM}^*)\\
\tau_T^* &= -\frac{16}{\pi}\sqrt{(v_{x'}^*)^2+(v_{y'}^*-\omega^*\ell_{CM}^*)^2}\left[C_L(\alpha)v_{x'}^*+C_D(\alpha)(v_{y'}^*-\omega^*\ell_{CM}^*)\right]\left[\ell_{CP}^*(\alpha)-\ell_{CM}^*\right]\\
\tau_{RL}^* &= -\frac{16}{\pi}C_R\omega^* v_{x'}^*(\ell_{CM}^*-\ell_{CRL}^*)\\
\tau_{RD}^* &=  -\frac{1}{4\pi}C_D^{\pi/2}\omega^*|\omega^*|\left[(2\ell_{CM}^*+1)^4 + (2\ell_{CM}^*-1)^4\right]\\
\tau_B^* &= - \frac{16}{\pi}(1-W^*)\ell_{CE}^*\cos\theta.
\end{split}
\end{equation}
Here $\ell_{CP}^* = \ell_{CP}/\ell$ and $\ell_{CM}^* = \ell_{CM}/\ell = W^* \ell_{CE}^*$. The Reynolds number $\Rey$ is the only dimensionless variable that does not appear explicitly. It should be viewed as implicit in the coefficients $C_R$, $C_L$, $C_D$, and $\ell_{CP}$ but which are modeled here as independent of $\Rey$ over the intermediate range of interest. Similarly, the aerodynamic effect of the slenderness ratio $h/\ell$ is included implicitly in these coefficients but which are modeled here as independent in the thin-plate limit $h/\ell \ll 1$. The static effect of $h/\ell$ due to weight and buoyancy is explicitly included via $W^*$.


\subsection{Survey of numerical solutions to the model}

\begin{figure}
\centering
    \includegraphics[scale=0.53]{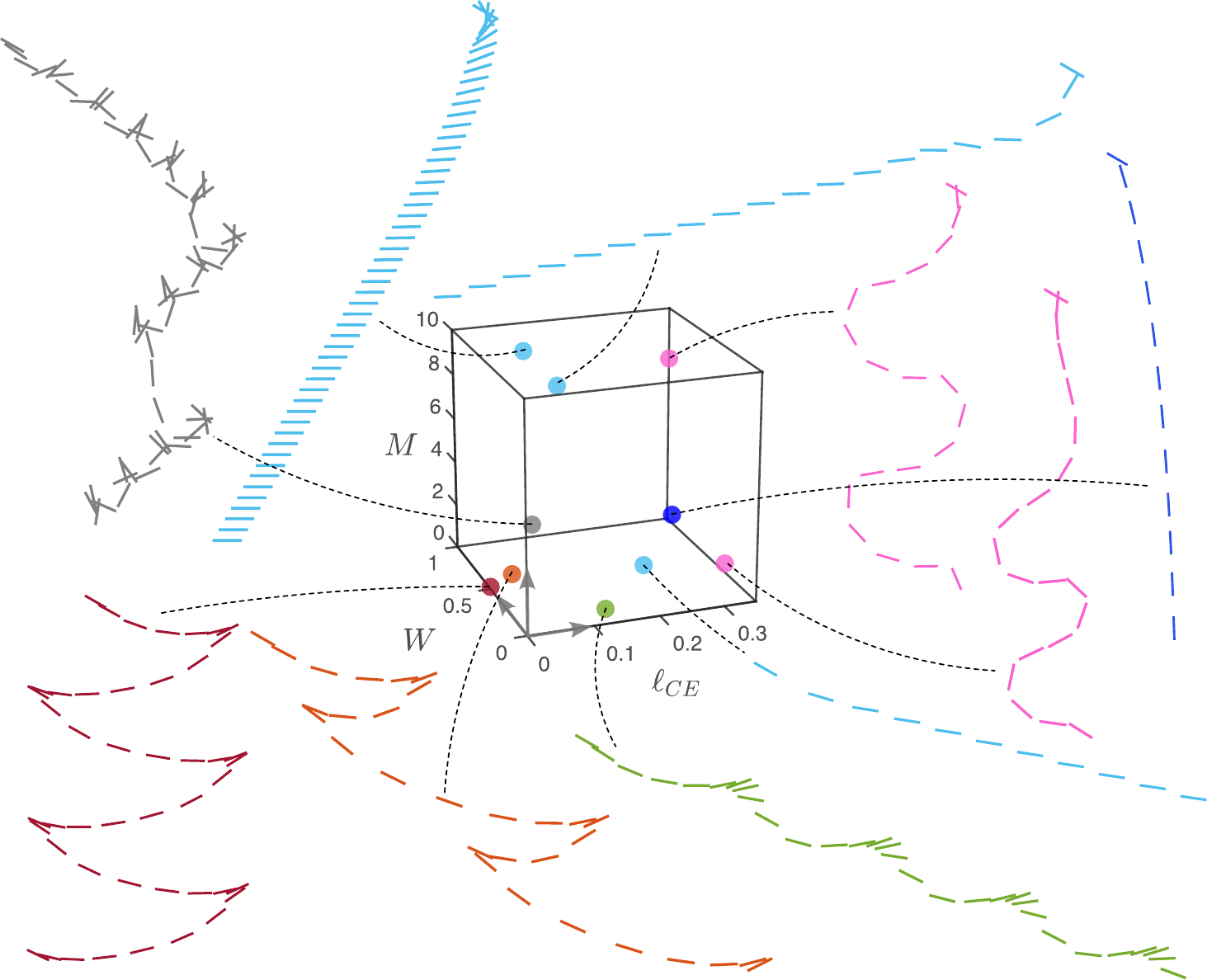}
  \caption{Sample trajectories produced by the flight dynamics model reveal a variety of behaviors. Different values of the parameters $(\ell_{CE},W,M)$ for fixed $I=0.1$ are marked on the 3D flight map, and the displayed plate motions result from identical initial conditions. Steady terminal states include gliding (light blue) at different attack angles and diving (dark blue) but pancaking is never observed. Periodic states include fluttering (red), progressive fluttering (orange), bounding (green) and meandering (pink). Aperiodic and apparently chaotic motions (grey) with bouts of tumbling are also observed.}
  \label{fig:trajectories}
\end{figure}

To give a sense of the types of motions produced by the model, we present in figure \ref{fig:trajectories} a variety of flight trajectories that arise as numerical solutions of the dynamical system for different parameter values. We hereafter work with the dimensionless system of equations \ref{ODEsystem_final} and \ref{lift_drag_torque} and drop the asterisks on $(\ell_{CE},W,M,I)$. These parameters serve as inputs to a ``flight simulator'' code that numerically integrates the ODEs in Matlab via the built-in solver \textit{ode15s}. The survey shown in figure \ref{fig:trajectories} explores the parameters $(\ell_{CE},W,M)$ for fixed $I = 0.1$, which is representative in that other values of $I$ produce similarly diverse sets of motions. Each case is marked by a point in parameter space, and the corresponding numerical solutions are displayed as snapshots of the plate orientation and location over time. All are released with the same initial conditions, and the resulting terminal or long-time motions vary greatly depending on the inputs. 

These samplings of the solution space may be familiar from fluttering leaves, erratically tumbling confetti, and flying paper planes. Some cases lead to steady motions such as sideways gliding (light blue) or downward diving (dark blue). Other cases lead to periodic behaviors such as back-and-forth fluttering with swoops punctuated by sharp reversals (red, orange), phugoid-like bounding with swoops of a single direction (green), or downward meandering along a smoothly winding course (pink). Others seem chaotic with bouts of end-over-end tumbling but lacking any repeated pattern (grey). The same flight pattern may be achieved for significantly different parameter values, and conversely a small change in parameters may lead to substantially different motions.

These explorations make clear the great complexity of the flight space encompassing how the governing parameters map to the eventual dynamical behavior. Importantly, a major result of this work is that one or more of the steady motions (gliding, pancaking and diving) exist as equilibrium solutions at any given point in parameter space. That such states emerge as solutions to the nonlinear model in some regions of parameter space and not others motivates the forthcoming stability analysis. More generally, the identification of equilibrium states and assessment of their stability will provide a way to systematically characterize the structure and organization of the flight space.

\section{Analytical forms of the free-flight equilibria}

The dynamical flight model admits analytical expressions for the equilibrium states. Here we continue working with the dimensionless system of equations \ref{ODEsystem_final} and \ref{lift_drag_torque} without the asterisks on the dimensionless parameters $(\ell_{CE},W,M,I)$. Recalling the reasoning of section 3, equilibria must have $\dot{v_{x'}}=\dot{v_{y'}}=\omega=0$ and so the nontrivial relations in equation \ref{ODEsystem_final} reduce to:
\begin{equation}\label{equilibrium_equations}
\begin{split}
0&=L_{x'}+D_{x'}-(2/\pi)\sin\theta\\
0&= L_{y'} + D_{y'} - (2/\pi)\cos\theta\\
0&= \tau_T +\tau_B.
\end{split}
\end{equation}
Since $M$ and $I$ do not appear in these relations, the claim from section 3 that neither parameter determines the equilibria is proven. That is, if an equilibrium is identified, it may be achieved for any value of either parameter. In contrast, $\ell_{CE}$ and $W$ appear in the $\tau_B$ term per equation \ref{lift_drag_torque}. Expanding the above equations via the definitions in equations \ref{lift_drag_torque} yields:
\begin{equation}\label{equilibrium_expanded}
    \begin{split}
    \sqrt{v_{x'}^2 + v_{y'}^2}\left[C_L(\alpha)v_{y'} -C_D(\alpha)v_{x'}\right] &= \sin\theta \\
        -\sqrt{v_{x'}^2 + v_{y'}^2}\left[C_L(\alpha)v_{x'} +C_D(\alpha)v_{y'}\right] &= \cos\theta \\
        \sqrt{v_{x'}^2+v_{y'}^2}\left[C_L(\alpha)v_{x'}+C_D(\alpha)v_{y'}\right]\left[\ell_{CP}(\alpha)-W\ell_{CE}\right] &= (W-1)\ell_{CE}\cos\theta.
    \end{split}
\end{equation}
In what follows, we simplify the notation by suppressing the functional dependencies on $\alpha$ for $C_L$, $C_D$ and $\ell_{CP}$. One can show by manipulating equations \ref{equilibrium_expanded} and invoking the four-quadrant inverse tangent definition of the dynamic attack angle $\alpha$ that solutions exist for the subset of values $\alpha\in [-90^\circ,90^\circ]\cup\lbrace -180^\circ\rbrace$, these corresponding to downward trajectories. Equivalently, steady-state solutions exist only for static attack angles $\alpha'\in[0^\circ,90^\circ]$. The general solution to all equilibria has the form: 
\begin{equation}\label{eq_solution}
    (v_{x'},v_{y'},\cos\theta,\ell_{CE}) = \left( \frac{\cos\alpha}{\left(C_L^2+C_D^2\right)^{1/4}},\frac{\sin\alpha}{\left(C_L^2+C_D^2\right)^{1/4}},-\frac{C_L\cos\alpha + C_D\sin\alpha}{\sqrt{C_L^2+C_D^2}},\ell_{CP}\right).
\end{equation}
The parameter $W$ does not appear, having dropped out of the equilibrium solution upon assigning $\ell_{CE}=\ell_{CP}$. This completes the proof of the claim in section 3 that $\alpha$ can be taken as the sole input that determines the equilibrium values of $\vec{v}^{CV}$, $\theta$ and $\ell_{CE}$, with $W$, $I$ and $M$ being free.

Gliding, diving and pancaking are specific instances within the general set of solutions. Gliding states are those solutions with $\alpha\in(-\pi/2,0)\cup(0,\pi/2)=(-90^\circ,0^\circ)\cup(0^\circ,90^\circ)$, or equivalently $\alpha'\in(0,\pi/2)=(0^\circ,90^\circ)$. These arise in pairs of leftward and rightward gliding of oppositely signed $\alpha = \pm \alpha'$ and orientation angles $\theta$ differing by $\pi = 180^\circ$. Diving states are those solutions with $\alpha=\lbrace 0,-\pi\rbrace=\lbrace 0^\circ,-180^\circ\rbrace$, corresponding respectively to bottom- and top-heavy postures, both of which have $\alpha'=0$. Manipulation of equations \ref{equilibrium_expanded} shows $\ell_{CE}$ to be a free variable rather than one constrained by $\ell_{CP}$. The solution pair for diving is: 
\begin{equation}
    (v_{x'},v_{y'},\theta,\ell_{CE}) = \left(\pm\frac{1}{\sqrt{C_D}},0,\mp\frac{\pi}{2}=\mp90^{\circ},\ell_{CE}\geq 0\right).
\end{equation}
Pancaking states are the pair of solutions with $\alpha=\pm\pi/2=\pm90^\circ$ or $\alpha'=\pi/2=90^\circ$: 
\begin{equation}
    (v_{x'},v_{y'},\theta,\ell_{CE}) = \left(0,\mp\frac{1}{\sqrt{C_D}},-\pi=-180^{\circ} ~\textrm{or}~ 0=0^{\circ},0\right),
\end{equation}
which are physically indistinguishable and thus degenerate.

\begin{figure}
\centering
    \includegraphics[scale=0.6]{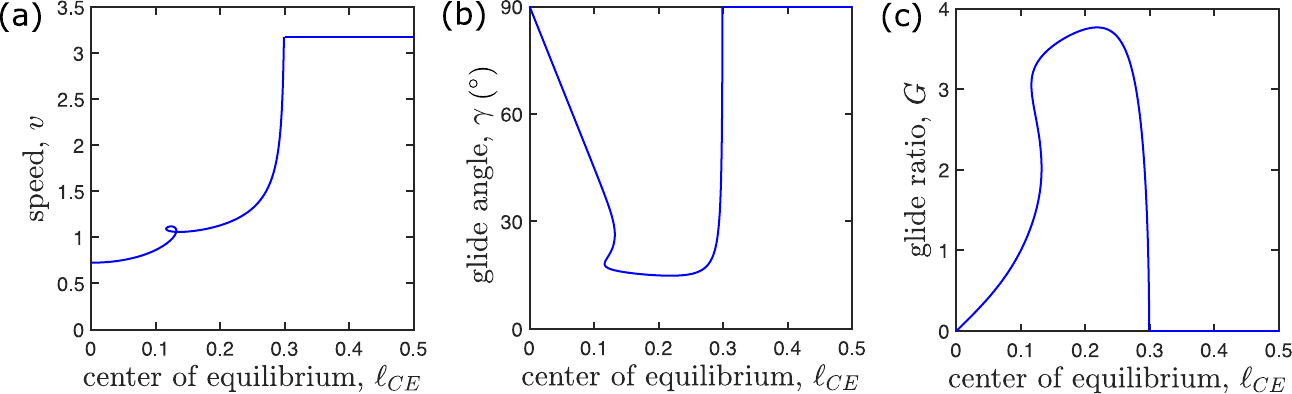}
  \caption{Lab frame quantities for equilibrium states as a function of static center of equilibrium. (a) Dimensionless flight speed achieved during pancaking ($\ell_{CE} = 0$), gliding ($\ell_{CE} \in (0,0.3)$), and diving ($\ell_{CE} \geq 0.3$). (b) Glide angle $\gamma$ measured relative to the horizontal. (c) Corresponding glide ratio $G$ representing the horizontal distance traveled per unit distance of fall.}  \label{fig:glide_angle}
\end{figure}

In the lab frame, the equilibrium solutions take the form of steady motion at constant speed $v$ along a straight trajectory of angle $\gamma$ relative to the horizontal:
\begin{equation}\label{glide_angle}
\begin{split}
    v = \frac{1}{\left(C_L^2+C_D^2\right)^{1/4}}, \quad \gamma = \alpha + \theta, \quad \textrm{and} \quad G = 1/\tan\gamma.
\end{split}
\end{equation}
Here the glide ratio $G$ is the horizontal distance traveled per unit vertical distance of fall. These dimensionless quantities depend on $\alpha$ and can therefore be recast in terms of $\ell_{CE}$, as shown in figures \ref{fig:glide_angle}(a)-(c). Pancaking ($\ell_{CE}=0$) and diving ($\ell_{CE}\geq 0.3$) involve direct downward descent and hence $\gamma = 90^\circ$ and $G=0$. Gliding states have $G>0$, and a maximal value of $G=3.8$ is seen near $\ell_{CE} = 0.22$, i.e., around the quarter-chord point \citep{li2022centre}.


In the following stability analyses, it will be sufficient to confine our attention to those states with $\alpha\in [0,\pi/2] = [0^\circ,90^\circ]$. That is, we need not concern ourselves with distinguishing the paired states, since they share the same stability. The exception is top-heavy diving with $\alpha=-\pi=-180^\circ$, which however can be dismissed as statically unstable for all parameters.

\section{Stability of gliding states}

With the dynamical system fully defined and its equilibrium solutions identified, we now seek to understand the parameter ranges for which the equilibria are stable. As such, we conduct a linear stability analysis. Due to the mathematical complexity of the model and its many terms, analytical derivations are available but extremely cumbersome and so we instead employ symbolic computational methods. We develop a code base written in MATLAB that (1) computes equilibria of the ODE system equations \ref{ODEsystem_final}, (2) symbolically linearizes the system about a given equilibrium state, and then (3) assesses the corresponding stability using eigenvalue analysis. Here we again work with the dimensionless variables $(\ell_{CE},W,M,I)$, where the asterisks have been dropped for convenience. The analysis presented here will cover gliding and pancaking states, i.e., $\alpha \in (0,\pi/2] = (0^\circ,90^\circ]$, with diving ($\alpha=0=0^\circ$ or $\pi=180^\circ$) to be handled separately.

The inputs to the algorithm are the equilibrium angle of attack $\alpha \in (0,\pi/2] = (0^\circ,90^\circ]$, the normalized and buoyancy-corrected weight $W\in (0,1)$, the relative mass $M>0$, and the relative moment of inertia $I>0$. The code first solves for the unique equilibrium state $(\ell_{CE}=\ell_{CP}(\alpha),\theta, \vec{v} = \vec{v}^{CM} = \vec{v}^{CV},\omega=0)$, which in all cases verify the analytical results of section 5. Then the Taylor expansion toolbox in MATLAB is used to symbolically linearize the dimensionless ODEs \ref{ODEsystem_final} about this equilibrium. The algorithm then computes the eigenvalues $\lambda_i$ of this linearized system, of which there are 4 in accordance with the 4 degrees of freedom $(v_{x'},v_{y'},\theta,\omega)$. Finally, the code assesses the stability of a particular equilibrium state by checking the sign of the real component of each eigenvalue. If the real components of the eigenvalues are all negative, then the equilibrium is classified as \textit{stable}. If there exists at least one eigenvalue with positive real component, then the equilibrium is unstable. Moreover, if any of these eigenvalues with positive real component also has a zero imaginary component, then the unstable equilibrium is called \textit{statically unstable}; otherwise, it is called \textit{dynamically unstable}. In summary, the progression is as follows:
\[
\begin{split}
(\alpha, W, M, I) \Rightarrow (\ell_{CE}, \theta, \vec{v}) \Rightarrow \lambda_i \Rightarrow \begin{cases}
    \text{stable} & \forall i, \text{Re}(\lambda_i) < 0 \\
    \text{statically unstable} & \exists i \text{ s.t. } \text{Re}(\lambda_i) > 0,~\text{Im}(\lambda_i) = 0 \\
    \text{dynamically unstable} & \text{otherwise}
\end{cases}
\end{split}
\]

\begin{figure}
\centering
    \includegraphics[scale=0.75]{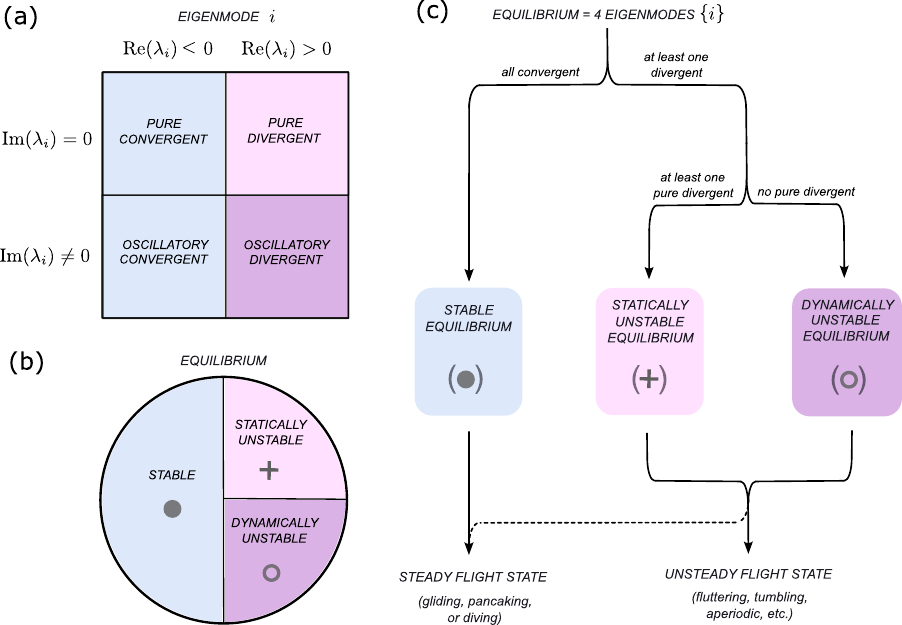}
  \caption{Classification schemes and nomenclature for eigenmodes and free-flight equilibria. (a) Each eigenmode $i$ may be one of four types based on the real and imaginary parts of its eigenvalue $\lambda_i$. (b) Each free-flight equilibrium may be one of three types: stable (blue), statically unstable (pink), or dynamically unstable (purple). (c) Flow diagram showing how the type of equilibrium is determined from the types of eigenmodes. Each equilibrium has 4 eigenmodes, and the presence or absence of certain mode types determines the stability status. } 
  \label{fig:venn_diagram}
\end{figure}

In summary, the 4-dimensional input space $(\alpha,W,M, I)$ is mapped to an output characterization that assumes one of the three possible classifications (stable, statically unstable, dynamically unstable), each of which corresponds to some steady or unsteady flight pattern. Since $\alpha \in (0,\pi/2] = (0,90^\circ]$ maps to a unique $\ell_{CE}$, we may also consider the analogous map from $(\ell_{CE},W,M, I)$ to (stable, statically unstable, dynamically unstable). This recasts all input parameters as the intrinsic physical properties of the plate-fluid system.

Terms such as static/dynamic stability/instability may be used differently and sometimes casually in other works \citep{etkin1995dynamics, anderson2005introduction, anderson2011ebook}. Our classification scheme is mathematically specific and complete and may therefore impart distinct and additional meaning. The schematics of figure \ref{fig:venn_diagram} relate the various terms. An \textit{equilibrium} is a set of dynamical parameters needed to specify a free-flight motion that balances forces and torques. Each equilibrium has 4 eigenvectors and associated eigenvalues that emerge from a stability analysis, and we call these modes or, more specifically, \textit{eigenmodes}. As shown in panel (a), each eigenmode may classified as one of 4 types with labels of \textit{convergent} or \textit{divergent} based on its real part and \textit{pure} or \textit{oscillatory} based on its imaginary part. As shown in (b), each equilibrium may be classified as being one of 3 types with names \textit{stable}, \textit{statically unstable}, or \textit{dynamically unstable}. As specified in (c), the presence or absence of certain types of eigenmodes determines the stability status of the equilibrium. Namely, a stable equilibrium has no divergent eigenmodes, a statically unstable equilibrium has at least one pure divergent eigenmode, and a dynamically unstable equilibrium has one or more divergent modes none of which are pure. Static instability is ``obvious'' in the sense that it would be apparent even in the setting of a wind tunnel where a single degree of freedom (e.g., attack angle) is statically perturbed and the measured response (pitch moment or torque) found to be destabilizing rather than restorative. In contrast, dynamic instability is the more subtle or ``hidden'' type that necessarily involves the free-flight couplings among multiple degrees of freedom. Note that the distinction between the two unstable equilibria is inconsequential in the sense that in either case the free-flight system veers away and displays some other dynamics.

These classifications and terms have some relations to free flight. We use the term free-flight state, flight state, or simply \textit{state} to describe the long-time or terminal behavior exhibited by the dynamical system for a set of parameters specifying an equilibrium. As shown in panel (c), a stable equilibrium will necessarily lead to a \textit{steady state} with the same characteristics (speed, attack angle, etc.) of the equilibrium, and so we call the state by same name of gliding, pancaking or diving used for the equilibrium. An unstable equilibrium of either type will typically lead to an \textit{unsteady state}, including periodic motions such as fluttering or tumbling and also aperiodic or non-repeating motions. However, the status of an equilibrium as statically or dynamically unstable tells us nothing of the nature of the free flight state beyond that it must be different from the equilibrium. Moreover, as indicated by the dashed route in (c), a given unstable equilibrium may in some cases exhibit a steady state whose dynamical parameters differ from the equilibrium. For example, an unstable diving equilibrium may lead to a stable gliding state as the terminal motion during free flight.

In what follows, we will chart out the stability of equilibria across wide ranges of the input parameters. Throughout these extensive investigations, we have spot-checked many cases by comparing the predictions of the linear analysis against numerical solutions of the full nonlinear model. The two always match. That is, conditions predicted to be linearly stable yield steady motion as the long-time numerical solution to the full model; dynamically unstable but statically stable equilibria yield purely oscillatory instabilities that manifest as growing fluctuations at short times; and statically unstable equilibria, which necessarily have at least one divergent mode and may also have an oscillatory mode or modes, veer away from the equilibrium continuously or with oscillations depending on the dominant eigenmode. This correspondence of the linear analysis with the short-time behavior of the nonlinear system near equilibria is expected for smooth (i.e., $C^{\infty})$ dynamical systems per the Hartman-Grobman theorem \citep{strogatz2018nonlinear, hartman1960, hartman2002ordinary}.

\subsection{Overview of the flight space}

\begin{figure}
\centering
    \includegraphics[scale=0.75]{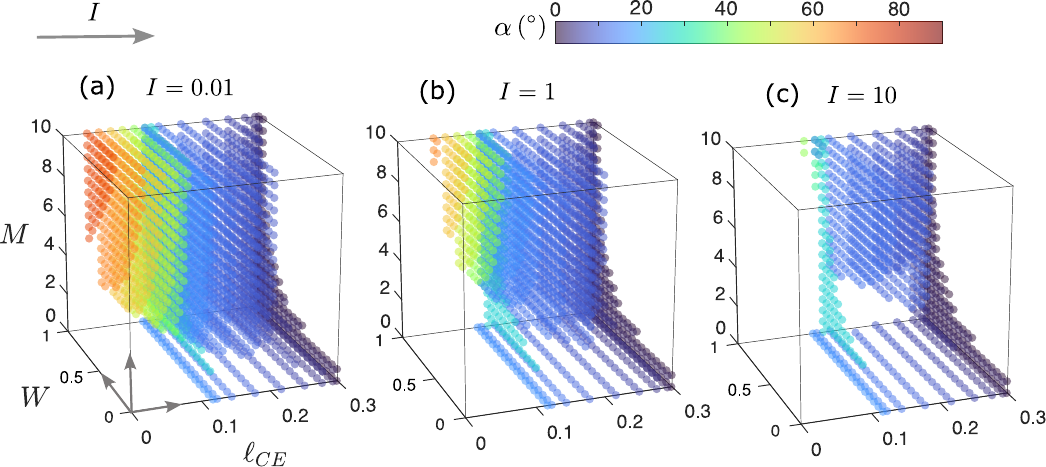}
  \caption{Overview of the 4D parameter space relevant to gliding and pancaking equilibria. Each plot shows the 3D space of the dimensionless parameters $(\ell_{CE},W,M)$ for fixed values of $I$: (a) $0.01$, (b) $1$, and (c) $10$. Each point in the space represents at least one equilibrium. Stable equilibria are indicated with markers colored by the attack angle $\alpha$ whereas unstable equilibria are left blank.}
  \label{fig:3d_gliding}
\end{figure}

The stability map of solutions across the 4-dimensional input parameter space $(\alpha, W, M, I)$ can be visualized by taking 3D sections in which one quantity is fixed. Figure \ref{fig:3d_gliding} shows 3D maps in terms of $(\ell_{CE}, W, M)$, where the corresponding $\ell_{CE}(\alpha)$ is displayed as an axis variable in place of $\alpha$ and where the moment of inertia $I$ is fixed within each panel and increases across the panels. The chosen values of $I = 0.01$, $1$ and $10$ are representative of low, intermediate, and high moments of inertia. Each point in the space corresponds to at least one equilibrium according to the preceding existence argument. Unstable equilibria are unmarked and thus correspond to the regions of white space in the plot boxes. Stable equilibria are marked with points whose color denote the angle of attack $\alpha$, as specified by the color bar. The axis range $\ell_{CE} \in [0,0.3]$ directly reflects the range for the $\ell_{CP}(\alpha)$ curve (figure \ref{fig:lift_drag}b) and the equilibrium condition $\ell_{CE}=\ell_{CP}$, and no gliding/pancaking states appear outside this range. The interval $W \in (0,1)$ is intrinsic to the problem of a dense object in a less dense medium. The range $M \in (0,10]$ is truncated at its upper bound only for convenience of plotting, and the solution space continues upwards for larger $M$ but with no significant change in structure. 

From the 3D sections of figure \ref{fig:3d_gliding}, it is clear that the available locations in parameter space for stable systems have a complex structure. Stable gliding may be achieved for greatly different values of any given parameter. There are significant and unique dependencies with respect to all parameters, indicating that all play essential roles in the stability. Surprisingly, the flight stability demonstrates independent behavior with respect to $W$ and $M$, these two respectively encoding the gravitational and inertial aspects of mass. The stable regions appear variously as 3D blobs and quasi-2D sheets, and these may be distinctly separated at places and bridged by quasi-1D filaments elsewhere. A tour through this space reveals some observations and trends that further emphasize this complexity:

\begin{itemize}
    \item Across the entire space, pancaking with $\alpha=90^\circ$ and thus $\ell_{CE}=0$ is always unstable. This corresponds to the white space seen in each panel of figure \ref{fig:3d_gliding} that covers the left coordinate plane. Systems in the near vicinity with low $\ell_{CE}$ also tend to be unstable, as indicated by the voids on the left sides of each panel. Falling plates thus have an aerodynamic aversion to broadside-on motions.
    
    \item For the low $I\ll 1$ case represented by figure \ref{fig:3d_gliding}a, stable gliding is realized across a wide range of $\alpha$, including high-$\alpha$ gliding for high $M$ and low $\ell_{CE}$ (red markers). Comparison across the panels shows that the availability of such motions decreases with increasing $I$.

    \item For the moderate $I=O(1)$ case represented by figure \ref{fig:3d_gliding}b, the central blob of stable solutions shrinks somewhat. Correspondingly, stable gliding is accessible only for low to moderate values of $\alpha$. This case also makes more clear the structure for low $M \ll 1$, for which stable gliding at low $\alpha$ is available for approximately $\ell_{CE} \in [0.1,0.3]$ and across all $W \in (0,1)$. This planar region of stability lies on the lower coordinate plane or floor of the space, and it is present for all $I$.

    \item For the high $I\gg 1$ case represented by figure \ref{fig:3d_gliding}c, the central blob shrinks further and stable solutions are confined to a yet lower range of $\alpha$. The low-$M$ stable region along the floor is yet more visible, and it is connected to the blob through a vertical filament that seems present for all $I$. Also more clear is a set of solutions of very small $\alpha$ (dark blue) that lies along the right wall or coordinate plane defined by $\ell_{CE} = 0.3$ and which exists for all $I$. 

\end{itemize}

\subsection{Two-dimensional dissections of the flight space}

\begin{figure}
\centering
    \includegraphics[scale=0.87]{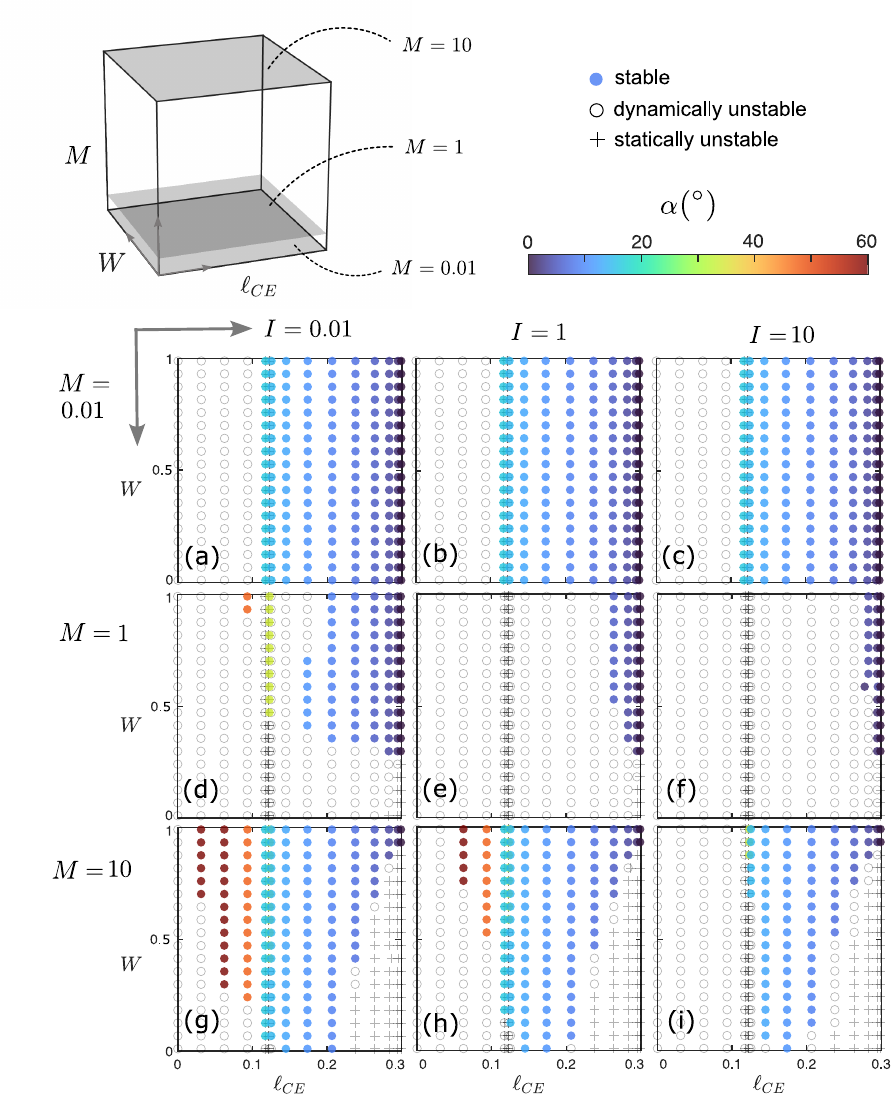}
  \caption{Matrix of 2D sections in the parameters $(\ell_{CE},W)$ for fixed $M$ and $I$ representing low, moderate and high values. The top left schematic shows the how the 2D sections relate to the 3D plots of figure \ref{fig:3d_gliding}. Stable states are shown with markers colored by the attack angle whereas dynamically and statically unstable equilibria are shown with grey $\circ$ and $+$ markers, respectively. The values of $(M,I)$ are: (a) $(0.01,0.01)$; (b) $(0.01,1)$; (c) $(0.01,10)$; (d) $(1,0.01)$; (e) $(1,1)$; (f) $(1,10)$; (g) $(10,0.01)$; (h) $(10,1)$; (i) $(10,10)$. }
  \label{fig:2d_gliding}
\end{figure}

\begin{figure}
\centering
    \includegraphics[scale=0.88]{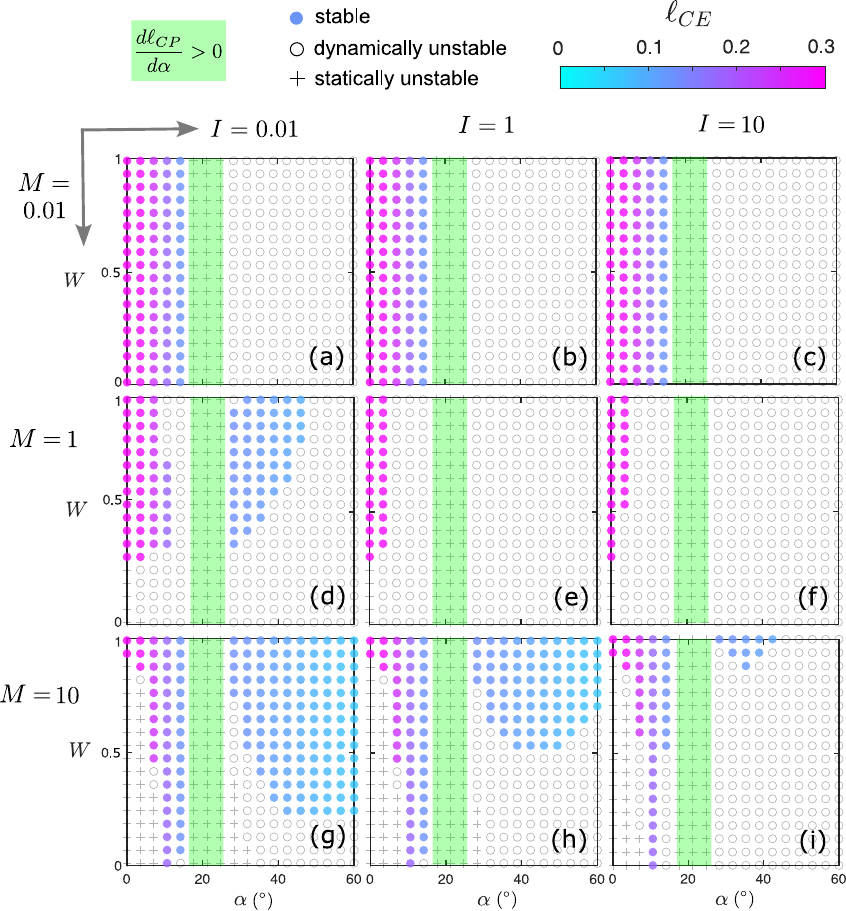}
  \caption{Matrix of 2D sections in the parameters $(\alpha,W)$ for fixed $M$ and $I$ representing low, moderate and high values. This information recasts that of figure \ref{fig:2d_gliding} so that $\alpha$ is an axis variable and the corresponding $\ell_{CE}=\ell_{CP}(\alpha)$ colors the stable states. Dynamically and statically unstable equilibria are again shown with grey $\circ$ and $+$ markers, respectively. Those equilibria with $d\ell_{CP}/d\alpha>0$ (green shading) are statically unstable. The values of $(M,I)$ are: (a) $(0.01,0.01)$; (b) $(0.01,1)$; (c) $(0.01,10)$; (d) $(1,0.01)$; (e) $(1,1)$; (f) $(1,10)$; (g) $(10,0.01)$; (h) $(10,1)$; (i) $(10,10)$.}
  \label{fig:2d_gliding_alpha}
\end{figure}

The space can be further dissected into 2D slices that are more easily displayed and examined. In the tableaux of figures \ref{fig:2d_gliding} and \ref{fig:2d_gliding_alpha}, we investigate the parameter space by taking several 2D slices at fixed $M$ and $I$. The values of $M=0.01,1,10$ are shown in increasing order downward through the panels, and the values $I=0.01,1,10$ increase across the panels. These sections are representative of low, moderate, and high cases in each variable, and there are no significant structural changes outside the displayed ranges. The miniature 3D diagram at the top of figure \ref{fig:2d_gliding} serves as a guide showing how the 2D slices relate to the 3D diagrams of figure \ref{fig:3d_gliding}. Figures \ref{fig:2d_gliding} and \ref{fig:2d_gliding_alpha} represent the same information, the former using $\ell_{CE}$ as an axis variable and $\alpha$ as the marker color for stable equilibria, and vice versa for the latter. Unstable equilibria of both types are shown as grey markers, where open circles represent dynamically unstable equilibria and crosses represent statically unstable equilibria. These collectively correspond to the white space in figure \ref{fig:3d_gliding}.

These findings reinforce many of the main messages from the 3D diagrams, notably that pancaking is always unstable while a rich variety of gliding motions are available for different combinations of parameter values. The 2D slices also provide deeper insights into the structure, dependencies and trends and especially additional information about the regimes for differing $M$:

\begin{itemize}
    \item Equilibrium solutions are unique with respect to 
    $\alpha$ (figure \ref{fig:2d_gliding_alpha}) but not unique with respect to $\ell_{CE}$ (figure \ref{fig:2d_gliding}), which is readily explained by the equilibrium condition $\ell_{CE}=\ell_{CP}$ and the nonmonotonic form of $\ell_{CP}(\alpha)$, as shown in figure \ref{fig:lift_drag}b. As such, any given location in the panels of figure \ref{fig:2d_gliding} may have overlapping points that represent multiple equilibria, some of which may be stable and others unstable. Such cases occur near $\ell_{CE} \approx 0.12$. Examples include panel (d) at high $W$, where stable gliding equilibria with $\alpha \approx 35^{\circ}$ (yellow points) appear together with dynamically and statically unstable equilibria (circles and crosses visible beneath the yellow points). At a similar location in panel (g), two stable gliding equilibria (blue and yellow points) overlap with statically unstable equilibria (crosses).

    \item For the low $M \ll 1$ case represented by panels (a-c) in figures \ref{fig:2d_gliding} and \ref{fig:2d_gliding_alpha}, stable gliding is generally available and its structure is simple. The maps are identical and thus independent of $I$, and the stability structure is also independent of $W$. The region over which stable gliding can be achieved is thus given by the approximate condition $\ell_{CE} \in (0.12,0.3)$ or equivalently, $\alpha \in (0^\circ,17^{\circ})$.

    \item For the moderate $M = O(1)$ case represented by panels (d-f), stable gliding equilibria are generally sparse and especially so for moderate and high $I$. For low $I$ as documented in panel (d), the stable regions appear as two islands separated by unstable equilibria.

    \item For the high $M \gg 1$ case represented by panels (g-i), stable gliding is generally available for low $I$ and less so with increasing $I$. The low-$\ell_{CE}$, high-$\alpha$ stable equilibria are increasingly cut off as $I$ increases.

\end{itemize}

\subsection{Role of center of pressure in gliding stability}

The complex structure of the 4D flight space reflects the fact that the eigenvalues and associated stability conditions lack simple mathematical formulas. Nonetheless, our characterizations reveal tidy relationships linking the center of pressure to stability that, while defying analytical derivations, are documented to exactly hold across the entire space of parameters. Specifically, we observe two relations involving the sign of the derivative of $\ell_{CP}(\alpha)$ or slope of the curve given in figure \ref{fig:lift_drag}b:
\begin{equation}
\textrm{stable gliding} \Rightarrow \frac{d\ell_{CP}}{d\alpha}< 0 \quad \textrm{and} \quad \frac{d\ell_{CP}}{d\alpha}> 0 \Rightarrow \textrm{statically unstable}
\end{equation}
The first assertion is that all stable gliding solutions have negative pressure center slope. The converse is not true, and there are many cases of negatively sloping $\ell_{CP}(\alpha)$ that are unstable. Hence, negative slope is a necessary but not sufficient condition for stability. The second assertion is that all equilibria associated with positive slope of the pressure center undergo static instability. The converse is again not true, and there are many statically unstable equilibria with negative slope. The second relation is visually confirmed by comparing figure \ref{fig:lift_drag}b, where negative slopes are seen for $\alpha \in (17^{\circ}, 26^{\circ})$, and figure \ref{fig:2d_gliding_alpha}, where the corresponding equilibria are highlighted in green and seen to be unanimously statically unstable ($+$ markers). Thus, gliding at any angle in this range is strictly forbidden as a free-flight stable state.

These observations connect to the flight dynamical notion of static (in)stability as pertaining to the torque response for a static perturbation to the attack angle in a wind tunnel setting \citep{etkin1995dynamics,anderson2011ebook,amin2019role}. Positive slope $d\ell_{CP}/d\alpha> 0$ means that the center of pressure moves forward on the wing for a nose-up change in $\alpha$, which is destabilizing. So too is the center moving back on the wing in response to a nose-down perturbation. Stable gliding therefore necessarily requires $d\ell_{CP}/d\alpha<0$, though this one condition alone is not sufficient to guarantee stability in free flight where all degrees of freedom participate.

\subsection{Numerical investigations into unsteady flight states resulting from unstable equilibria}

The analysis presented above identifies vast regions of parameter space where flight is unstable, which correspond to the voids in figure \ref{fig:3d_gliding} and grey markers in figures \ref{fig:2d_gliding} and \ref{fig:2d_gliding_alpha}. It is for these cases that the full nonlinear model yields unsteady flight motions as its numerical solutions. In this section, we extend the survey of such motions presented in section 4.3 and figure \ref{fig:trajectories} and delve deeper into cases of interest. We focus on several 2D sections of the flight space and show that our model successfully reproduces states observed in previous studies and also predicts new free-flight patterns in heretofore unexplored regions of the parameter space.

\begin{figure}
\centering
    \includegraphics[scale=0.37]{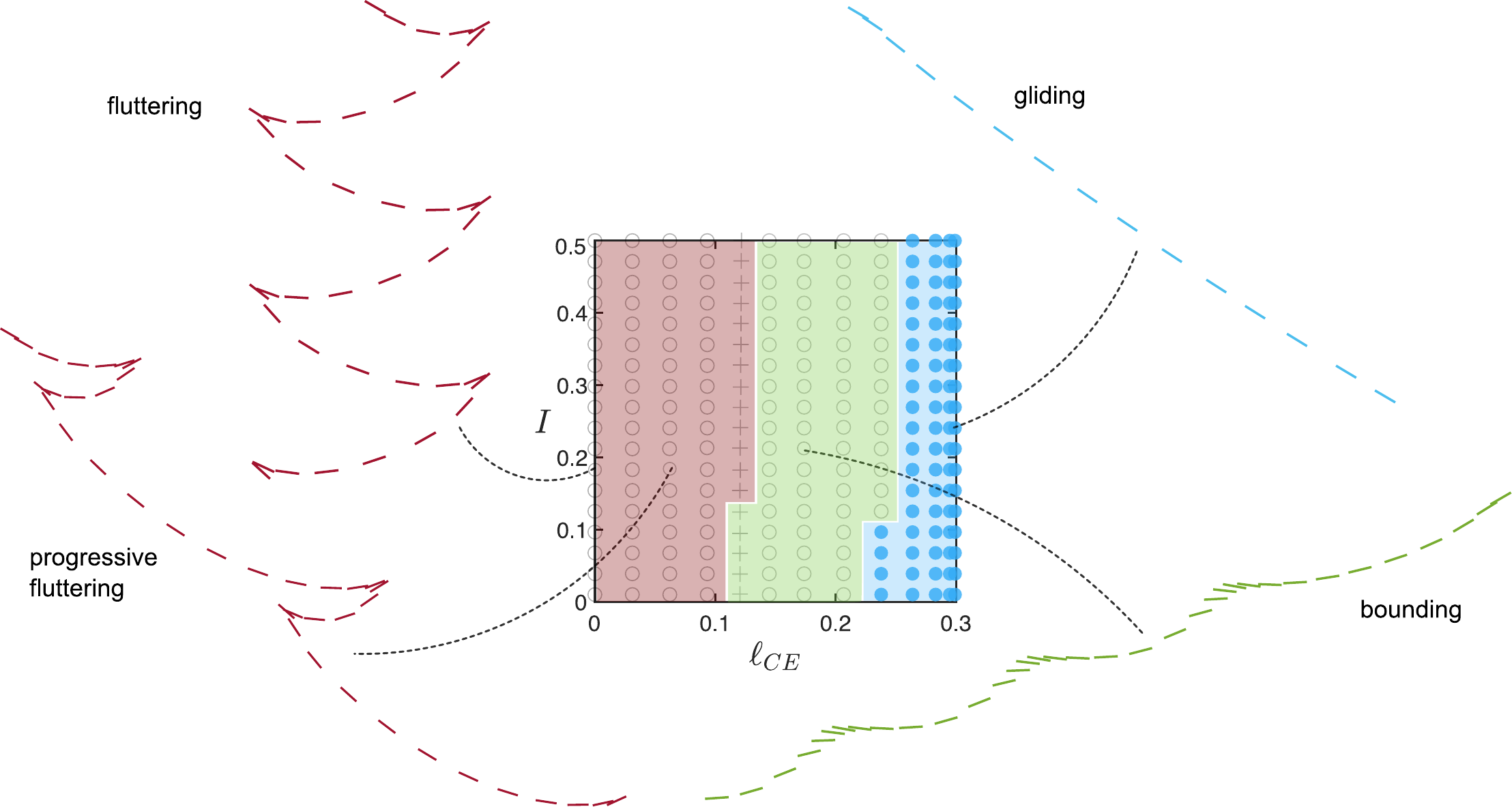}
  \caption{Validation of the model against the experiments of \citet{li2022centre} whose conditions correspond to varying $(\ell_{CE},I)$ for fixed $(W,M)=(0.2,0.14)$. The four experimentally observed states of fluttering, progressive fluttering, bounding and gliding are reproduced by the model, whose output plate dynamics are shown. The shading delineates the state regimes determined by simulation runs across the 2D parameter space.}
  \label{fig:tiffany_slice}
\end{figure}

A first case involves comparisons to the modeling and experimental results of \citet{li2022centre} on the flight behaviors displayed by plates with displaced centers of mass. In particular, this earlier work reported on fixed values of $M=0.14$ and $W=0.2$, for which different unsteady flight states were identified for differing values of $\ell_{CE}$ and $I$. The modifications in our model relative to that of \citet{li2022centre} prove to be inconsequential with respect to these results, all of which are faithfully reproduced. In figure \ref{fig:tiffany_slice}, we display the relevant region in a 2D slice of parameter space with coloring that denotes 4 unsteady states: fluttering with symmetric back-and-forth swoops, progressive fluttering with asymmetric swoops, bounding with one-way swoops, and steady gliding. The 4 displayed plate motions are numerical solutions to our nonlinear model corresponding to the parameters explored in \citet{li2022centre}. These results confirm the stability type in all cases. Further explorations show these states to be generally available, as indicated by shaded regions identified by running simulations across the ranges $\ell_{CE} \in [0,0.3]$ and $I \in (0,0.5]$. The state boundaries are largely dictated by the former and have weaker dependence on the latter.

\begin{figure}
\centering
    \includegraphics[scale=0.37]{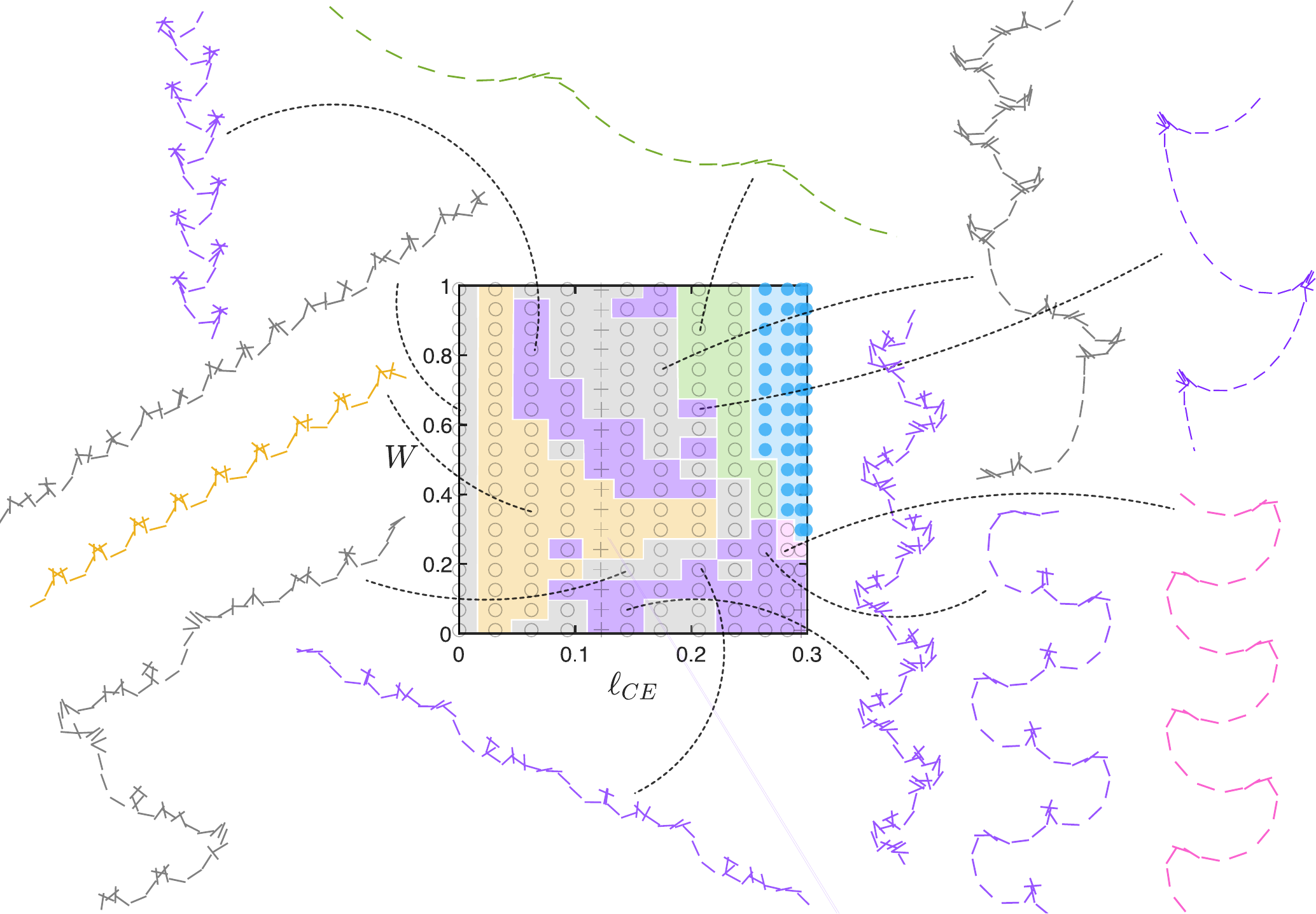}
  \caption{Complex and varied motions dominate for intermediate values of the inertial parameters $(M,I)=(1,1)$ where gliding is significantly depleted. This panel repeats figure \ref{fig:2d_gliding}e and adds shading whose different colors indicate the flight states observed in simulation runs across the space. The motions include familiar states such as gliding (blue), bounding (green), meandering (pink) and tumbling (yellow) but also many other periodic states (purple) of varied forms as well as aperiodic and apparently chaotic states (grey). }
  \label{fig:chaos}
\end{figure}

\begin{figure}
\centering
    \includegraphics[scale=0.32]{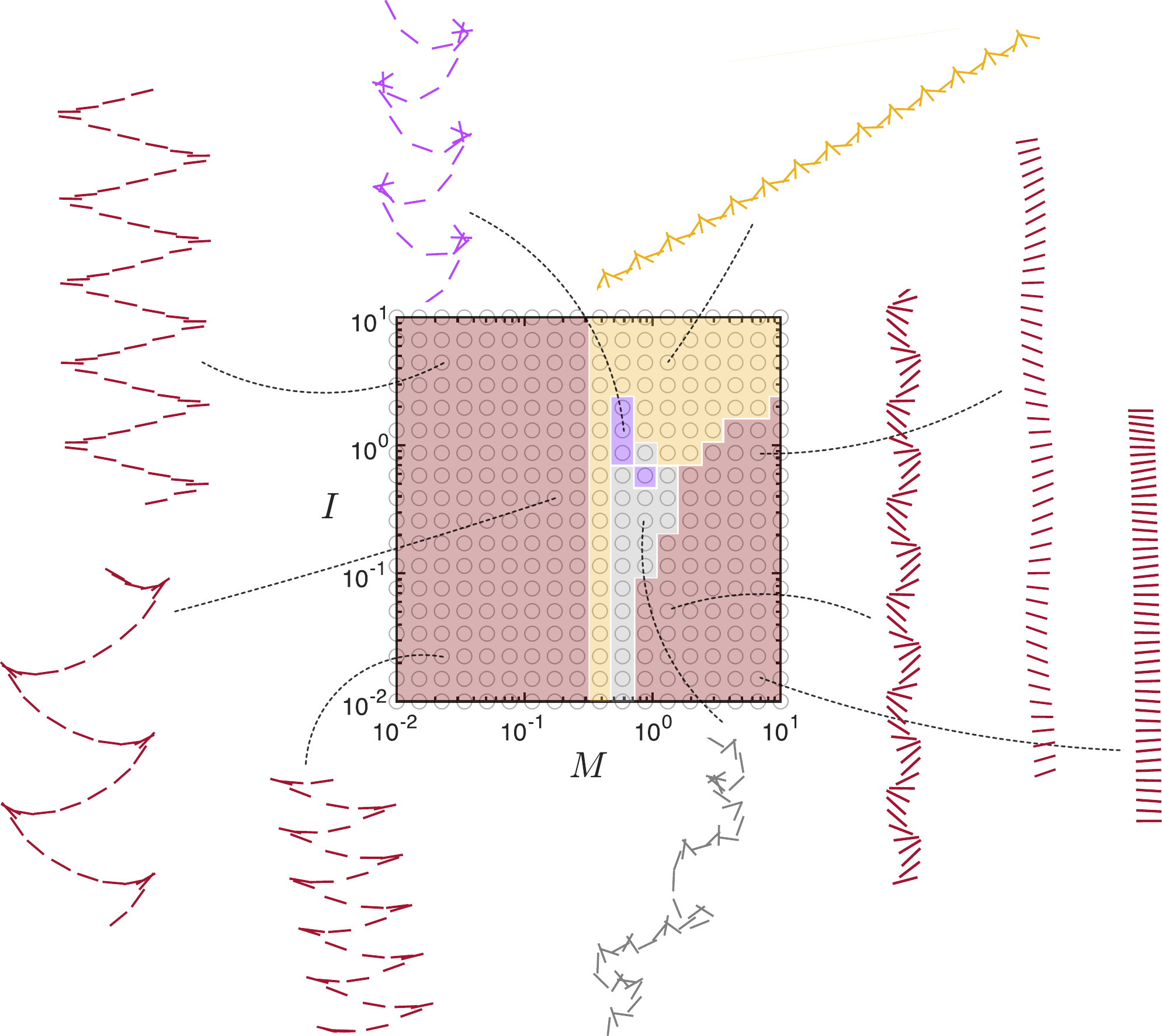}
  \caption{Pancaking is dynamically unstable ($\circ$) for all conditions. For fixed $(\ell_{CE},W)=(0,0.8)$, the parameters $(M,I)$ are varied logarithmically to explore the space broadly. The shading indicates the flight state observed in simulations, and example trajectories are displayed. The motions are predominantly forms of fluttering (red) and tumbling (yellow) but with additional periodic (purple) and aperiodic (grey) states occurring for intermediate $M$ and low to intermediate $I$.}
  \label{fig:broadside_2}
\end{figure}

Bounding (shown in green in figure \ref{fig:tiffany_slice}) is a limit-cycle state that superficially resembles the phugoid motion of aircraft \citep{montalvo2015meta} but which differs in that the angle of attack increases significantly in what look like approaches to stalling. Such behavior was previously documented for plates with displaced center of mass \citep{li2022centre}. Bounding occurs for parameters near the stable-unstable border with gliding, and this structure is consistent with a Hopf bifurcation \citep{guckenheimer2013nonlinear}.

Irregular and aperiodic motions which involve elements of familiar tumbling and fluttering motions are reported by \citet{andersen2005analysis} and \citet{hu2014motion} for moderate $M=O(1)$ and $I=O(1)$. Our model reproduces these motions in this intermediate inertial regime, where gliding availability is significantly depleted and a significant majority of equilibria are unstable. In figure \ref{fig:chaos}, whose map is that of figure \ref{fig:2d_gliding}e which has $(M,I)=(1,1)$, we systematically chart the numerical solutions to the nonlinear model for each of the unstable equilibria by shading the state space according to associated long-term flight patterns. The regions are blotchy and irregular, with complex boundaries and islands demarcating the various states. In grey we shade aperiodic and apparently chaotic states, which display no discernible pattern over any timescale; in yellow we shade tumbling; in pink we shade meandering, which is a heretofore unreported flight state described in greater detail in section 7.4; in green we shade bounding; and in light blue we shade stable gliding. In purple we shade those regions which admit periodic hybrid motions that combine multiple characteristics of tumbling, fluttering, meandering, and bounding. Regions of these irregular motions appear in a highly complex fashion in the state space, and more intricate boundaries may exist at a finer resolution of the space than is explored here.

Pancaking or broadside-on falling of a symmetric plate has been reported by \citet{andersen2005analysis}, who found the motion unstable for any $I$ under the $\Rey$ conditions assumed in this work. We report that indeed, pancaking is dynamically unstable for any set of parameters. In figure \ref{fig:broadside_2}, we hold $W=0.8$ and $\ell_{CE}=0$ fixed and systematically chart the numerical solutions to the nonlinear model for each of the unstable equilibria by shading the $(M,I)$ state space according to associated long-term flight patterns. All equilibria are open $\circ$ markers, and so pancaking is always statically stable but dynamically unstable. This figure is representative of any choice of $W$, since by equations \ref{ODEsystem_final}, $W$ drops out for $\ell_{CE}=0$. We plot the information here on a logarithmic scale to better interrogate the intermediate inertial regime where irregular motions abound. In red we shade those regions which display fluttering, and show several different types of fluttering ranging from very weak (right) to very strong (left). As in figure \ref{fig:chaos}, we shade tumbling, aperiodic states, and periodic hybrid states using yellow, grey, and purple, respectively.



\section{Stability of diving states}

We separately assess diving states, which involve edgewise downward descent as depicted in figure \ref{fig:gliding_diving}b. Recalling the arguments of section 3 and the analytical expressions for equilibria of section 5, there are two possible diving postures: a bottom-heavy configuration with the center of equilibrium displaced towards the leading edge side ($\alpha=0=0^\circ$) and a top-heavy configuration with $\ell_{CE} > 0$ displaced towards the trailing edge ($\alpha=-\pi=-180^\circ$). As expected intuitively, the latter is statically unstable for all parameter values, so throughout this section we focus on the former. We may treat all cases of interest $\ell_{CE} \geq 0$, including the symmetric plate of $\ell_{CE} = 0$ which is degenerate with respect to the two diving postures. As we have shown, diving is distinguished from gliding in that $\ell_{CE}$ is free since the center of pressure constraint is inoperative. We therefore modify the MATLAB code for stability analysis of the ODE system \ref{ODEsystem_final} by supplying $\ell_{CE}$ as an input for fixed $\alpha=0$. We otherwise use the same classification scheme as in the gliding case:
\[
\begin{split}
(\ell_{CE},W,M,I) \Rightarrow (\theta, \vec{v}) \Rightarrow \lambda_i \Rightarrow \begin{cases}
    \text{stable} & \forall i, \text{Re}(\lambda_i) < 0 \\
    \text{statically unstable} & \exists i \text{ s.t. } \text{Re}(\lambda_i) > 0,~\text{Im}(\lambda_i) = 0 \\
    \text{dynamically unstable} & \text{otherwise}
\end{cases}
\end{split}
\]
This yields a map from the 4-dimensional input space $(\ell_{CE},W,M,I)$ to an output characterization that assumes one of the three stability classifications. The following investigations into diving stability parallel those for gliding.

\subsection{Overview of the flight space}

\begin{figure}
\centering
    \includegraphics[scale=0.45]{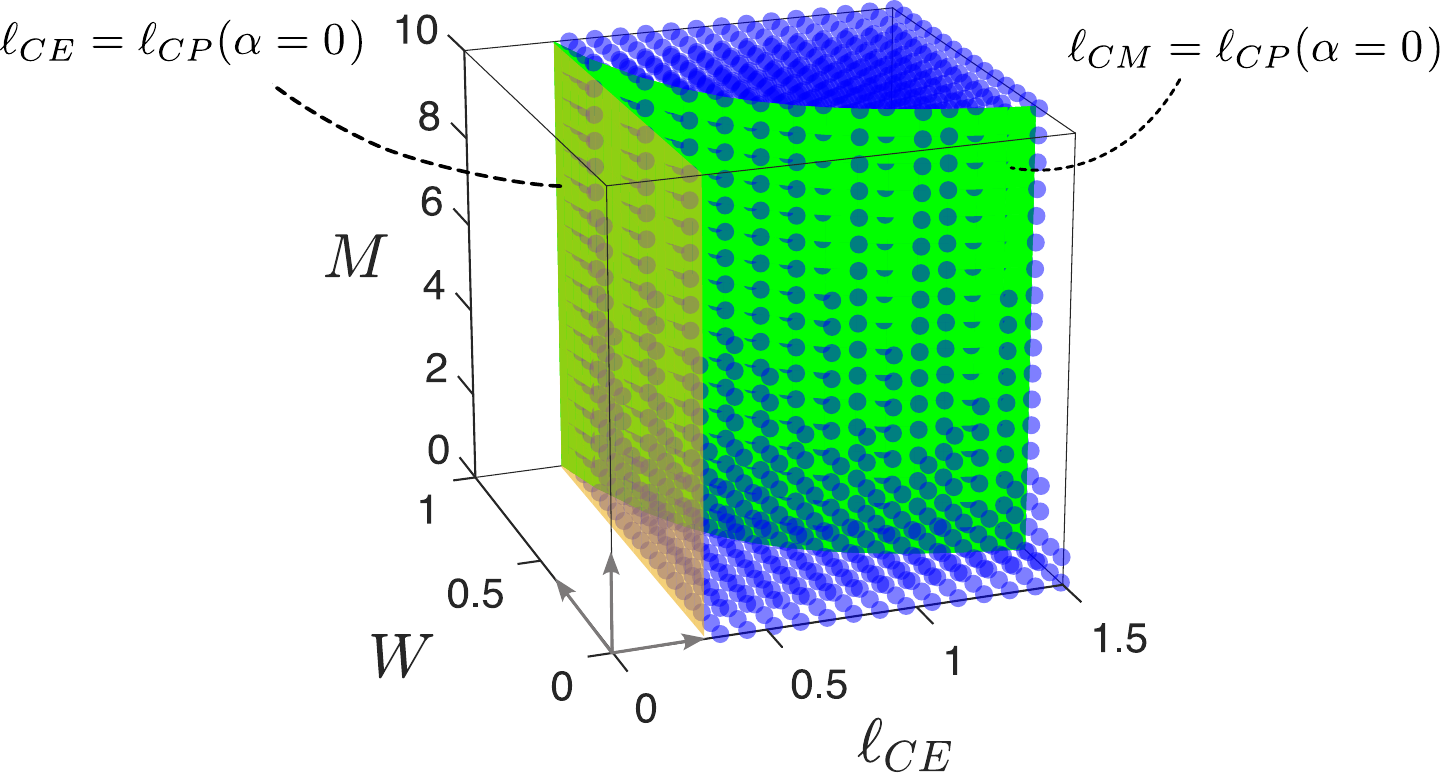}
  \caption{Diving stability ascribes to a simple structure in the parameter space, shown here across varying $(\ell_{CE},W,M)$ for a representative value of $I=1$. Each point represents a unique equilibrium that is either stable (blue markers) or unstable (blank). The surfaces $\ell_{CE}=\ell_{CP}(\alpha=0)=0.3$ (orange) and $\ell_{CM} = W\ell_{CE} = \ell_{CP}(\alpha=0)=0.3$ (green) demarcate stability boundaries in different regimes of $M$. }  \label{fig:3d_diving}
\end{figure}

The 4-dimensional input parameter space $(\ell_{CE},W,M,I)$ for diving may be visualized by taking 3D sections in which one quantity is fixed. Figure \ref{fig:3d_diving} shows such a 3D space of $(\ell_{CE},W,M)$ with the moment of inertia $I=1$ fixed at a representative intermediate value. Stable equilibria are marked with colored points, which are uniformly blue to indicate that $\alpha = 0$ for all. Each such point corresponds to exactly one equilibrium according to the preceding existence and uniqueness arguments. Unstable equilibria are unmarked and thus correspond to the blank spaces in the plot box.

The displayed axis range is $\ell_{CE}\in[0,1.5]$, and there is no significant change in structure for yet greater values. Note that this range extends beyond that for gliding, reflecting that $\ell_{CE}$ is a free parameter no longer constrained by $\ell_{CP}\in[0,0.3]$. As for gliding, the interval $W\in(0,1)$ is intrinsic to the problem. The range $M\in(0,10]$ is truncated at its upper bound only for the convenience of plotting, and there is no change in structure for yet greater values. Also shown are two surfaces $\ell_{CE}=\ell_{CP}(\alpha=0)=0.3$ (orange upright plane) and $\ell_{CM}=W \ell_{CE}=\ell_{CP}(\alpha=0)=0.3$ (green hyperbolic sheet) which correspond to some stability boundaries, as discussed further below.

From figure \ref{fig:3d_diving}, it is clear that the available regions of parameter space for stable diving solutions have a simple structure composed of two distinct but connected regions: 
\begin{itemize}
    \item For low $M < 1$, there exists a quasi-planar or sheet-like region where diving is stable for all $\ell_{CE}>0.3$ regardless of $W$. This forms the floor of the flight space, and it is bounded on one side by the orange surface.
    \item For high $M > 1$, there exists a 3D region bounded by the curved upright wall (green surface). Here, stable diving involves conditions on both $\ell_{CE}$ and $W$.
\end{itemize}

\subsection{Two-dimensional dissections of the flight space}

\begin{figure}
\centering
    \includegraphics[scale=0.9]{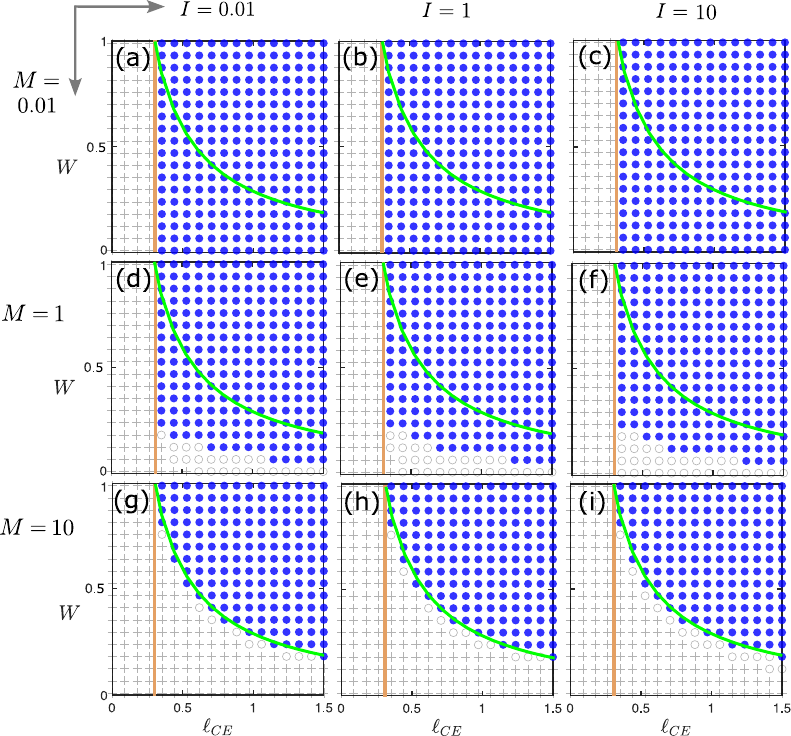}
  \caption{Matrix of 2D sections in the parameters $(\ell_{CE},W)$ for fixed $M$ and $I$ representing low, moderate and high values. The middle column with $I=1$ represents constant-$M$ sections of figure \ref{fig:3d_diving}.
  Stable diving states are shown as blue markers whereas dynamically and statically unstable equilibria are shown with grey $\circ$ and $+$ markers, respectively. The orange and green curves are sections through the corresponding surfaces of figure \ref{fig:3d_diving} and which are important stability boundaries in the limits of low and high mass. The values of $(M,I)$ are: (a) $(0.01,0.01)$; (b) $(0.01,1)$; (c) $(0.01,10)$; (d) $(1,0.01)$; (e) $(1,1)$; (f) $(1,10)$; (g) $(10,0.01)$; (h) $(10,1)$; (i) $(10,10)$. }   \label{fig:2d_diving}
\end{figure}

More refined insights come from taking 2D sections of the flight space. In the tableau of figure \ref{fig:2d_diving}, we show slices of $(\ell_{CE},W)$ across representative low, moderate, and high values of $M$ and $I$. Here the plot markers again indicate stability status with blue filled markers for stable equilibria, crosses for statically unstable equilibria, and open circles for dynamically unstable equilibria. The latter two unstable cases correspond to the white space in figure \ref{fig:3d_diving}. These data reinforce the messages distilled from the 3D space and add more details:
\begin{itemize}
    \item There is exceedingly little variation with the moment of inertia $I$. Changes with $I$ are not altogether absent but they are few and limited to the borders representing transitions between stability and instability or between static and dynamic instabilities. As examples, one may closely compare panels (d) and (e) in the vicinity of $(\ell_{CE},W)=(0.4,0.1)$.
    \item Sufficiently low $\ell_{CE} < \ell_{CP}(\alpha=0)=0.3$ is always statically unstable across all values of $(W,M,I)$. Note that, strictly speaking, the center of pressure is undefined at $\alpha=0$ (figure \ref{fig:lift_drag}b) and hence should be understood here in the limiting sense. This result means that the plate must be sufficiently front weighted to have a chance of being stable in diving. The relevant boundary is the orange vertical line in each panel, which corresponds to the section through the orange upright plane in figure \ref{fig:3d_diving}.
    \item For low mass represented by $M = 0.1$ in panels (a-c), the simple condition $\ell_{CE} > \ell_{CP}(\alpha=0)=0.3$ seems necessary and sufficient for stable diving. This boundary is shown in all panels as the orange vertical line.
    \item For moderate mass represented by $M = 1$ in panels (d-f), $\ell_{CE}$ must be yet greater to ensure stability when $W$ is low. Those unstable solutions with $\ell_{CE} > \ell_{CP}(\alpha=0)=0.3$ are mostly but not exclusively of the static-stable but dynamic-unstable type, meaning they will destabilize via growing oscillations.
    \item For high mass represented by $M = 10$ in panels (g-i), the simple condition $\ell_{CM} = W \ell_{CE} > \ell_{CP}(\alpha=0)=0.3$ seems necessary and sufficient for stable diving. This boundary is shown as the green hyperbolic curve in each panel, which is the section through the green hyperbolic surface in figure \ref{fig:3d_diving}. For high mass, dynamically unstable equilibria tend to hug closely the boundary, and the unstable equilibria are otherwise of the static type.
\end{itemize}

Regarding the near independence of stability status on $I$, we lack an explanation for this fact. The mathematical expressions for stability conditions do indeed contain this parameter but it apparently has exceedingly weak effect. Perhaps analysis of the expressions for the eigenvalues could give insight. Regarding the transition in stability status with $M$, we discuss some interpretations in what follows.

\subsection{The roles of the centers of pressure, equilibrium, and mass in diving stability}

The above results can be summarized by the following conditions for dynamic stability:
\begin{equation}
\begin{split}
\textrm{stable diving at low $M \ll 1$} &\Longleftrightarrow \ell_{CE} > \ell_{CP}(\alpha=0)=0.3 \quad \textrm{and} \\ \textrm{stable diving at high $M \gg 1$} &\Longleftrightarrow \ell_{CM} = W \ell_{CE} > \ell_{CP}(\alpha=0)=0.3
\end{split}
\end{equation}
The two-way arrows indicate necessary and sufficient relations. In words, low-mass stable diving requires the center of equilibrium to be forward of the center of pressure, and high-mass stable diving requires the center of mass to be forward of the center of pressure. These outcomes are empirically validated but we lack proofs, and so the above claims should be viewed as conjectures that are empirically validated here across wide ranging conditions. Future work may pursue derivations by analyzing the eigenvalues in the appropriate limits.

Intuitively, these conditions state that plates must be weighted sufficiently forward for stable diving. This seems to relate to the directional or yaw stability of aircraft, which presumably have $M \gg 1$. For this so-called weathervane effect, the center of mass is viewed as a pivot point that must be sufficiently forward of the vertical stabilizer on the tail where pressure forces act \citep{anderson2005introduction}, which is consistent with the second condition given above. Our findings are also quantitatively consistent with the experiments of \citet{li2022centre} involving flight tests in water of plastic plates to which weights were added to the displace the centers of mass and equilibrium. Here the low relative mass $M=0.14$ means the first condition applies, and indeed the observation by \citet{li2022centre} that diving occurs only for $\ell_{CE}>0.3$ aligns with our findings. 

Less intuitive is that different regimes of $M$ involve different specifications of what exactly constitutes sufficient front weighting, one case involving $\ell_{CE}$ and the other $\ell_{CM}$. This indicates that the two factors play distinct roles, defying the intuition that the former is simply a generalization of the latter in situations in which buoyancy is a factor. Similarly, the inertial and gravitational aspects of mass play different roles, with $M$ dictating which condition is applicable and $W$ appearing within one of the conditional statements. Better understanding the origin of these subtleties is an avenue for future work. Our preliminary investigations show that the newly added torque from rotational lift $\tau_{RL}$ is in general an important determinant for diving stability. Without such a term, the model reverts to that of \citet{li2022centre}, whose stability maps differ substantially in the regime of high $M$. Namely, the stable region is then $\ell_{CE} \in (\ell_{CP},\ell_{CP}/W)$, which lies between the orange and green curves in the sections of figure \ref{fig:2d_diving}g-i. This seems implausible in that strongly front-weighted plates are predicted to be unstable. Future experiments would provide valuable information that would distinguish the models and perhaps further inform on the role of $\tau_{RL}$ and its mathematical form.

\subsection{Meandering: a new unsteady flight state}

\begin{figure}
\centering
    \includegraphics[scale=.8]{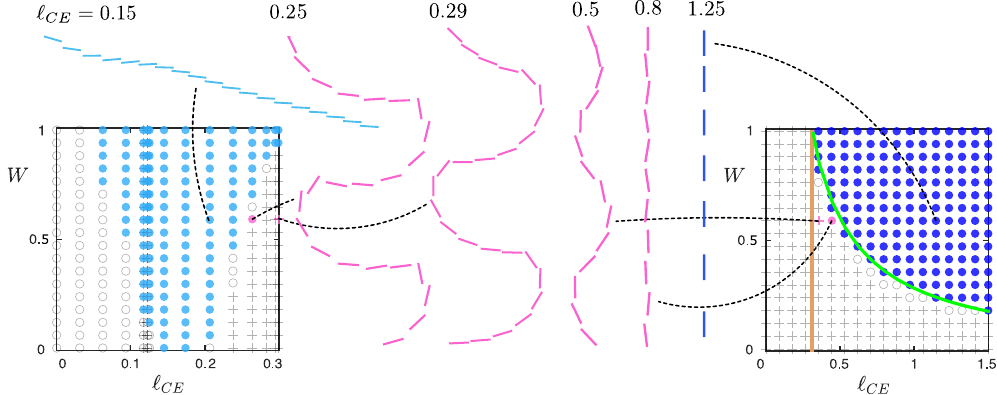}
  \caption{Meandering is a periodic state arising from unstable equilibria between stable gliding and stable diving. The panels represent fixed $(M,I)=(10,1)$ and varying $(\ell_{CE},W)$ over ranges appropriate to gliding and diving, and they repeat those of figures \ref{fig:2d_gliding_alpha}h and \ref{fig:2d_diving}h. Sample meandering trajectories (pink) of varying horizontal excursion amplitudes are accessed by varying $\ell_{CE}$ across values between stable gliding (light blue) and stable diving (dark blue). }  \label{fig:meandering}
\end{figure}

Our investigations into the stability of diving predict an unsteady state which has not to our knowledge been documented in previous studies. We call such behavior \textit{meandering}, which consists of smoothly waving side-to-side excursions during descent. Example trajectories are shown in pink in figures \ref{fig:trajectories} and \ref{fig:chaos}. The main characteristics of this new class of motions include:

\begin{itemize}
    \item Meandering, like fluttering, involves periodic back-and-forth oscillations about a directly downward trajectory. Whereas fluttering has sharply cusped reversals during which the plate does not flip over, meandering involves smooth turns during which the plate inverts in the sense of which face is projected upwards. Mathematically, the distinction is that $\hat{y'}\cdot\hat{y}$ is single-signed for fluttering but alternates in sign for meandering. 
    \item Meandering can originate as an instability of low-$\alpha$ gliding. Examples are shown on the left of figure \ref{fig:meandering}, for which the map repeats that of figure \ref{fig:2d_gliding}h with $M=10$ and $I=1$. The light blue points are uniformly colored here to indicate stable gliding regardless of attack angle, and a corresponding stable gliding trajectory is shown ($\ell_{CE}=0.15$, light blue). Meandering states (pink) arise as $\ell_{CE}$ is increased, and this appearance of a limit cycle state near a stability border is consistent with a supercritical Hopf bifurcation \citep{guckenheimer2013nonlinear}. The side-to-side excursion amplitude for meandering is highest for those states nearest the gliding stability boundary, i.e., for $\ell_{CE}$ closer to the plate center.    
    \item Meandering can also arise as an instability of diving, which parallels how fluttering emerges in a similar way from pancaking. Examples are shown on the right of figure \ref{fig:meandering}, for which the map repeats that of figure \ref{fig:2d_diving}h with $M=10$ and $I=1$. Here a stable diving trajectory is shown in blue ($\ell_{CE}=1.25$). Meandering states (pink) are found at the lower values of $\ell_{CE}$ corresponding to the region between the orange and green curves. Here again the stability structure is reminiscent of a supercritical Hopf bifurcation, and the excursion amplitude is seen to decrease with increasing $\ell_{CE}$.
    \item The results of figure \ref{fig:meandering} are typical of $M\gg 1$, for which meandering generally arises for those $\ell_{CE}$ too large to admit stable gliding and yet not sufficiently large to achieve stable diving. This solution can result from the region of instability regardless of static stability ($+$ markers) or static instability ($\circ$). The decreasing amplitude of meandering for increasing $\ell_{CE}$ can be understood intuitively: cases of low $\ell_{CE}$ represent failures to glide involving large lateral excursions, whereas cases of higher $\ell_{CE}$ are failures to dive with small excursions. 
\end{itemize}

Meandering may have been missed in previous studies because the requisite conditions were not explored. The work of \citet{li2022centre} seems to be the only experiments and modeling to address the necessarily large displacements in center of equilibrium $\ell_{CE} \gtrsim 0.2$, but the mass $M\approx 0.1$ was limited to low values for which figure \ref{fig:2d_diving}a shows the necessary unstable equilibria for meandering to be absent. Future experiments or numerical simulations testing our predictions could use figure \ref{fig:meandering} as a guide in choosing the relevant parameter values. For example, the case of $(\ell_{CE},W,M,I)=(0.25,0.6,10,1)$ yields the high-amplitude meandering, and such conditions should be achievable with strongly front-weighted plates falling in water.

\section{Discussion and conclusions}

This study provides a flight dynamics model for thin plates moving under gravity at moderate Reynolds numbers and methods for identifying equilibrium flight states and assessing their aerodynamic stability. Our approach builds on a quasi-steady aerodynamic framework that has been iteratively improved over recent decades to account for wider classes of flight modes including unsteady motions such as fluttering and tumbling and also steady motions such as gliding and diving. We conduct a dimensional analysis and assessment of equilibrium states to identify a minimal set of physical parameters that must be specified to uniquely determine the available steady motion, which is shown to be one of edgewise diving, broadside-on pancaking, or gliding. We then systematically investigate a dimensionless form of the model via linear stability analysis to reveal a complex set of conditions for which such a mode is stable as a free-flight state. Pancaking is predicted to be universally unstable whereas stable gliding and diving can be reached through distinct combinations of the center-of-equilibrium location $\ell_{CE}$, buoyancy-corrected weight $W$, mass $M$, and moment of inertia $I$. These findings represent a wealth of predictions that can be tested in future experiments and direct numerical simulations. The unstable regions of parameter space correspond to dynamic flight states, including a new class of motions called meandering that are shown to arise as limit cycle solutions for parameters between stable gliding and stable diving. Another useful consequence is the identification of a physical inconsistency in previous treatments of the rotational lift effect, for which we propose a remedy.

The model presented here involves many simplifying assumptions and idealized treatments of the aerodynamic effects and hence much room for improvement. The hope is that the large set of predictions presented here can help to guide future experimental and/or computational investigations that may identify shortcomings and thus drive refinement of the model. More generally, the analyses presented here may prove useful as means of systematically characterizing the results of future versions of this model or other such quasi-steady formulations. The following is a list of some key assumptions that could be better informed by future works as well as new predictions to be tested:

\begin{itemize}
    
    \item The model is quasi-steady and thus cannot address any mode transitions or changes in stability that are driven dominantly by wake dynamics, vortex shedding, or other intrinsically unsteady flow phenomena. Such effects are known to be important for $\Rey < 100$ \citep{ern2012wake, assemat2012onset, tchoufag2014global}, and hence the framework is necessarily limited to higher Reynolds numbers.

    \item Also related to unsteadiness, experiments and/or simulations should test the prediction of stable gliding for plates of high mass $M$ at high attack angles $\alpha$. One expects fully separated flow and an unsteady wake that may undermine stability in ways not captured by a quasi-steady framework.

    \item Our model follows previous versions in assuming that the aerodynamic coefficients are independent of Reynolds number. This should be taken as a low-order approximation and suitable modifications should be made if future work shows $\Rey$-dependent effects to be important for mode or stability transitions.

    \item More specifically on the above point, we take the skin friction coefficient $C_D^0$ to be a constant, which was shown in previous works to be sufficient for capturing known transitions \citep{andersen2005analysis, andersen2005unsteady, li2022centre}. The next-order approximation could involve a $\Rey$-dependent form based on Blasius boundary-layer theory \citep{schlichting2016boundary, bhati2018role}.

    \item The model further assumes that the aerodynamic coefficients are independent of the plate thickness, an approximation appropriate in the thin-plate limit of slenderness ratio $h/\ell \ll 1$. The work of \citet{li2022centre} showed that such a model yields good agreement with experiments for $h/\ell = 0.001$ to $0.1$, and accounting for thicker plates would likely require slenderness-dependent coefficients.
    
    \item The newly added term $\tau_{RL}$ representing the torque from rotational lift employs a force center $\ell_{CRL}=0$. This value is taken for simplicity, and future experiments or simulations would be informative.
    
    \item The newly predicted meandering state should be tested in lab experiments and/or direct numerical simulations conducted for the conditions identified here. The existence and form of this mode may depend on $\tau_{RL}$ and hence such tests could inform on this term.
    
\end{itemize}


While our model strictly applies to flat plates at intermediate Reynolds numbers as appropriate to the falling paper problem, its general form may be applicable more broadly to thin wings. Airfoil shapes, for example, would surely have different aerodynamic coefficients, and an appropriately modified model could be tested in future work. If validated, such models would have a broad range of potential uses for a variety of physical and biological systems. The computational tools for analysis introduced here could similarly be used to efficiently explore the large parameter space, as a pen-and-paper analysis is not feasible for such a complex set of equations. A major success of this work is that a variety of motions, both steady and unsteady, are predicted across widely ranging parameters. This suggests promise for the many applications involving motions in air and through water, both in understanding biological locomotion as well as for designing biomimetic flying and swimming vehicles.

\backsection[Acknowledgments]{We thank H. Li, C. Mavroyiakoumou, Z. J. Wang and J. Wu for useful discussions.}

\backsection[Funding]{We acknowledge support from the U.S. National Science Foundation (DMS-1847955).}

\backsection[Declaration of interests]{The authors report no conflict of interest.}


\backsection[Author ORCIDs]{O. Pomerenk, https://orcid.org/0000-0002-0481-3952; L. Ristroph, https://orcid.org/0000-0001-9358-0689.}

\backsection[Author contributions]{O.P. and L.R. contributed to all aspects of this work.}

\FloatBarrier

\appendix

\section{}\label{appA}

\begin{table}
\centering
\begin{tabular}{|c c c c c c c c c c|} 
 System & $M$ & $m$ & $s$ & $\ell$ & $L$ & $R$ & $h$ & $\rho_f$ & Reference \\
 \hline
  Bird & 691 & 62.1 & 68 & 33.5 & 40 & 5.5 & 1 & 1.2e-3 & \citet{berg1995moment}\\
  Watercraft & 60000 & 20000 & 120 & 20 & 140 & 12 & 1 & 1 & \citet{wood2009autonomous} \\
   Paper in air & 0 & 0.59 & 15 & 5.1 & 0 & 0 & 0.01 & 1.2e-3 & \citet{li2022centre}\\
 Plastic plate in water & 0 & 9.12 & 20.3 & 2.54 & 0 & 0 &0.15 & 1 & \citet{li2022centre}\\
 Stingray & 0 & 8000 & 60 & 50  & 0 & 0 &2& 1 & \citet{yigin2012age}\\
 Metal plate in water & 0 & 8.31 & 19 & 1 & 0 & 0 & 0.162 & 1 & \citet{andersen2005unsteady}\\
 Flying squirrel & 0 & 70 & 11 & 13 &0 & 0 & 3 & 1.2e-3 & \citet{thorington1981body}\\
Cucumber seed & 0 & 0.21 & 12 & 6.2 & 0 & 0 & 0.1 & 1.2e-3 & \citet{viola2022flying}\\ 
Butterfly & 0 & 2 & 20 & 20 & 0 & 0 & 0.005 & 1.2e-3 & \citet{hu2010experimental}\\
Snowflake & 0 & 3.3e-5 & 0.4 & 0.4 & 0 & 0 & 0.001 & 1.2e-3 & \citet{langleben1954terminal}\\
Marine snow & 0 & 0.0001 & 0.05 & 0.05 & 0 & 0 & 0.001 & 1 & \citet{passow2012marine}\\
Scallop & 0 & 100 & 6 & 6 & 0 & 0 & 1 & 1 & \citet{cheng1996dynamics}\\
Flounder & 0 & 2300 & 30 & 60 & 0 & 0 & 1 & 1 & \citet{takagi2010functional}\\
Air vehicle & 0 & 0.265 & 10 & 1.5 & 0 & 0 & 0.1 & 1.2e-3 & \citet{wood2007autonomous}
\end{tabular}
\caption{Systems that may be approximated as winged cylindrical bodies. All units are $\text{cm}\cdot\text{g}$. Displayed are $M$, mass of the fuselage body; $m$, mass of the wing; $s$, span length of the plate; $\ell$, chord length of the wing; $L$, length of the fuselage body; $R$, radius of the fuselage body; $h$, thickness of the wing; and $\rho_f$, density of the fluid.}\label{app_table}
\end{table}
For the order-1 calculations to produce figure \ref{fig:system}, we approximate fliers as systems which have a cylindrical fuselage body of mass $M$, length $L$, and radius $R$, with a thin rectangular wing of thickness $h$, chord length $\ell$, span length $s$, and mass $m$ cutting through the center of the cylinder (a simple ``airplane" structure). The ambient fluid has density $\rho_f$. Bodies which may be approximated as just wings, e.g., a piece of paper, have $M=L=R=0$. To calculate dimensionless quantities for this winged cylindrical body system, we compute using table \ref{app_table} the quantities
\begin{equation}
    \begin{split}
        I_{\text{body}} &= \frac{1}{12}M(3R^2+L^2) \\
        I_{\text{wing}} &= \frac{1}{12}m(\ell^2+h^2) \\
        I_{\text{wing (air)}} &= \frac{1}{2}\rho_f\pi s \left(\frac{\ell}{2}\right)^4
    \end{split}
\end{equation}
where the wing-air moment of inertia is a cylindrical fluid column with $I=\frac{1}{2}mr^2$. Then, we compute
\begin{equation}
    \begin{split}
        I^* &= \frac{I_{\text{body}} + I_{\text{wing}}}{I_{\text{wing (air)}}}\\
        M^* &= \frac{M+m}{\pi\rho_f s \left(\frac{\ell}{2}\right)^2} \\
        W &= g(M+m) \\
        B &= g\rho_f(s\ell h + \pi R^2L) \\
        W^* &= \frac{W-B}{W}.
    \end{split}
\end{equation}

\bibliographystyle{jfm}
\bibliography{jfm}

\end{document}